# Recoverable and plastic strains generated by forward and reverse martensitic transformations under external stress in NiTi SMA wires

*P. Šittner, E. Iaparova, O. Molnárová, O. Tyc, X. Bian, L. Kadeřávek, L. Heller*

Institute of Physics of the Czech Academy of Sciences, Na Slovance 2, Prague, Czech Republic

## Abstract

Superelastic NiTi #1 shape memory alloy (SMA) wire was subjected to thermomechanical loading tests in tension to evaluate stress and temperature conditions under which the wire deforms plastically. Although the alloy demonstrated a very high resistance to plastic deformation in both the austenite and martensite phases, incremental plastic strains were recorded whenever the B2 cubic to B19' monoclinic martensitic transformation (MT) took place under external stress. These incremental plastic strains, along with internal stresses and permanent lattice defects generated during the MT under stress accumulated over cyclic thermomechanical loading causing functional fatigue.

To shed light on the origin of the functional fatigue presumably originating from the accumulated plastic strains, special closed-loop thermomechanical loading tests were performed to evaluate recoverable and plastic strains generated by the forward and reverse MTs under external stress. These experiments revealed that both forward and reverse MTs generate well-defined incremental plastic strains only if they occur above certain stress thresholds. Specifically, the forward MT upon cooling under stresses up to 500 MPa does not produce plastic strain or lattice defects, whereas the reverse MT upon heating generates plastic strains even under stress as low as 100 MPa. Given that the reverse MT generates significantly larger plastic strains in thermal cycles than the forward MT, it is proposed to be largely responsible for functional fatigue of NiTi actuators.

It is proposed that plastic deformation generated by martensitic transformations under low stresses takes place by [100](001) dislocation slip in martensite and while under high stresses it occurs via kwinking deformation in martensite. The characteristic thresholds and magnitudes of plastic strains generated by the forward and reverse MTs define the functional fatigue limits for NiTi. Specifically, the stress-strain-temperature responses recorded in closed-loop thermomechanical tests remain stable if both forward and reverse MTs occur below these thresholds. Conversely, they become unstable and exhibit predictable ratcheting if the MTs exceed these thresholds.

# 1. Introduction

The stress-strain-temperature thermomechanical responses of NiTi shape memory alloys (SMA) due to B2 to B19' martensitic transformation [1] should be phase and strain reversible in closed loop thermomechanical load cycles, provided that the austenite and martensite phases do not deform plastically. This is, however, observed only when the MT proceeds at very low applied stresses. Whenever the martensitic transformation takes place under elevated external stress (Figs. 1c,d) plastic strains are generated [2-10]. Residual strains [2,3], internal stress [4], and lattice defects [5-9] accumulate during cyclic thermomechanical loading, leading to instability in cyclic thermomechanical responses, known as "functional fatigue" [11]. Despite decades of theoretical and experimental investigation, the precise mechanism by which martensitic transformation generates plastic strains and lattice defects remains unclear.

We have recently analyzed plastic strains and lattice defects in austenite generated by the forward and reverse MTs in superelastic nanocrystalline NiTi wire subjected to various thermomechanical load cycles (10 closed loop thermomechanical cycles) by post-mortem transmission electron microscopy (TEM) [5]. Our findings revealed that the accumulated plastic strains and lattice defects differ between the forward and reverse MTs and depend critically on the temperature and stress conditions under which the MTs occurred. However, our focus was on the accumulation of plastic strain and lattice defects over various cyclic thermomechanical loads, we did not evaluate the plastic strains and lattice defects generated by a single MT. We later recognized the urgent need to understand the mechanisms by which plastic strains and lattice defects are generated during single MTs.

Various theoretical concepts were proposed in the literature concerning the plastic strains generated by stress induced MT in NiTi. Some authors argue that dislocation defects accumulating during superelastic cycling of NiTi were created via dislocation slip in austenite at the habit plane interface [2,12,13,14]. They consider the martensite created by the stress-induced MT to be <011> type II twinned, a configuration frequently observed experimentally [15] and predicted by the Phenomenological Theory of Martensite Crystallography (PTMC) [16]. However, stress-induced martensite in NiTi containing (001) compound twins has also been reported experimentally [18-20,22-24].

Understanding stress-induced MT in NiTi requires knowledge of the habit plane interface that ensures strain compatibility across the interface. Matsumoto et al. [21] analyzed habit plane interfaces of stress induced MT observed in experiments on NiTi single crystals. They reported habit planes on the (-0.8684, 0.26878, 0.4138) austenite plane, which closely align with the theoretical PTMC solution considering <01-1> type-II twinning as lattice invariant shear (LIS). According to this widely cited result, stress-induced B19' martensite in NiTi is commonly considered to be <011> type-II twinned and connected to the austenite via the mentioned habit plane interface, despite numerous experimental reports showing habit plane interfaces between austenite and detwinned martensite [18,19].

Strain compatibility is automatically fulfilled at habit planes predicted by PTMC theory [21], which move during the forward and reverse MTs. Consequently, there should be no need for dislocation slip to assist such stress-induced MT in NiTi. Despite this, incremental plastic strain accompanying MT under stress is frequently observed experimentally [2,3,5-9,13,14]. To explain this, it was proposed that slip dislocations are generated during the MTs under stress, nucleating at structural heterogeneities of the moving habit plane interface, specifically at the steps formed by the martensite twins at the habit plane interface [2,13,14]. However, the relationship between this mechanism and the plastic strains generated by the martensitic transformation under external stress remains unclear.

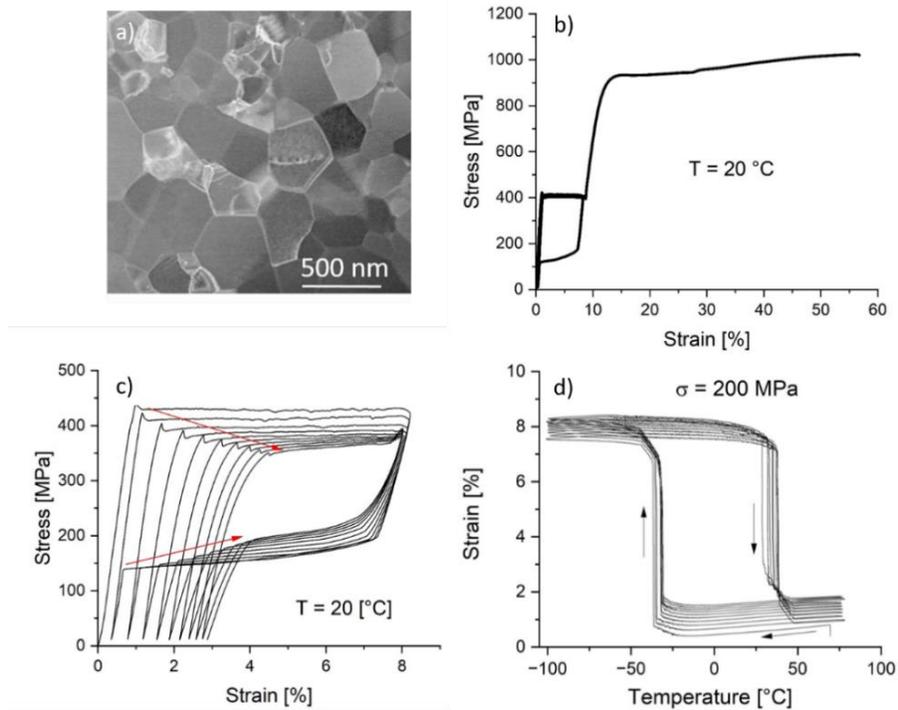

**Figure 1: Basic thermomechanical characteristics of the superelastic 15 ms NiTi #1 wire.** a) virgin austenitic microstructure (grain size 250 nm), b) stress-strain curve from tensile test at room temperature (one superelastic cycle until the end of the stress plateau followed by deformation until fracture, c) 10 superelastic cycles until the end of the stress plateau at room temperature, d) 10 thermal cycles under 200 MPa tensile stress documenting cyclic instability of stress-strain-temperature response of the wire due to plastic strains generated by the forward and reverse MTs proceeding under stress.

Many investigators have recently reported stress-induced martensite in NiTi, which is detwinned and/or (001) compound twinned [18-28]. In such cases, strain compatibility at the habit plane interface, moving during the martensitic transformation, is not automatically fulfilled but must be achieved somehow. Solving this strain compatibility problem is crucial for understanding MT under external stress. Cayron [29,30] proposed that strain compatibility at the habit plane between austenite and detwinned martensite can be achieved by considering plastic deformation via dislocation slip in austenite (an updated PTMC theory incorporating local plastic strains due to dislocation slip in austenite [16]). A different solution, considering anisotropic elastic strains of austenite and martensite, was recently proposed by Heller and Sittner [31]. They evaluated the magnitudes and orientations of critical uniaxial stress in tension and compression, for which strain-compatible habit plane interfaces between austenite and a single variant of martensite can form in NiTi. This calculated habit plane interface is not solely determined by the lattice parameters and lattice correspondence between austenite and martensite, but it varies with the magnitude and orientation of the applied stress since elastic strains also matter. Simulation results [31] explain the large spread of experimentally determined habit planes of stress-induced MTs observed decades ago (Fig. 6 in [32]). Nevertheless, since many experiments show plastic strain and a high density of slip dislocations in austenite generated by the martensitic transformation under external stress, it is likely that plastic strains neglected in [31] are somehow involved in ensuring strain compatibility at habit plane interfaces in NiTi.

When analyzing the unstable stress-strain-temperature responses of NiTi wires in cyclic isothermal (Fig. 1c) or isostress (Fig. 1d) thermomechanical loading tests, several intriguing questions arise regarding the physical origin of cyclic instability: i) whether the incremental plastic strains and permanent lattice defects are generated by the forward and/or reverse MTs, ii) whether plastic deformation occurs in the austenite and/or in the martensite and iii) why the incremental plastic strains are repeatedly generated whenever the MT proceeds under stress and what controls the magnitudes of incremental plastic strains.

It is worth mentioning that plastic strains and lattice defects generated by the martensitic transformation have been predominantly investigated in the literature following the reverse MT in closed-loop thermomechanical load cycles [5,6,12,18,19]. However, the observation of slip dislocations in austenite does not necessarily imply that they were created via plastic deformation of austenite. They could have originated in the martensite and been inherited by the austenite during the reverse MT upon unloading and/or heating. Plastic deformation of the low symmetry monoclinic martensite is, however, less understood [33-38] than plastic deformation of cubic austenite.

Plastic strains generated separately by single forward and single reverse MTs can be evaluated through closed-loop thermomechanical loading experiments, originally introduced by Heller et al. [10] and later utilized to investigate plastic strains and lattice defects in austenite or martensite generated by forward and/or reverse MT under stress [5,10,27,28]. Plastic strains generated by the forward and/or reverse MT under stress increase with the increasing stress under which they occur [5,10,20]. Notably, the reverse MT proceeding under stress was found to generate larger plastic strains than the forward MT (see Fig. 9 in Ref. [10], Fig. 15 in Ref. [5]). While these findings were intriguing, the experiments reported in Refs. [5,10,20]

were insufficient to reveal the mechanisms by which the forward and reverse MT generate plastic strains. We lacked information on martensite variant microstructures, textures, habit planes, and lattice defects created by single forward and reverse MTs occurring under stress. Therefore, further research was conducted in this direction [8,27,28], the results of which are briefly presented and utilized in the discussion in this work.

The aim of the research presented in this work is to uncover the mechanisms by which the forward and reverse martensitic transformations under stress generate plastic strains and consequently lead to functional fatigue of NiTi. To achieve this goal, we evaluated the plastic strains generated by single forward and reverse MTs occurring under a wide range of applied stresses in closed-loop thermomechanical loading tests. These experiments were conducted on superelastic NiTi#1 wire with optimized nanocrystalline microstructure and desired functional properties. The mechanisms by which the forward and reverse MTs generate plastic strains are proposed based on the obtained results as well as on the results of closely related studies, including i) the analysis of permanent lattice defects in austenite [8], and ii) the analysis of martensite variant microstructure and textures in austenite and martensite evolving during stress-induced MT during isothermal tensile loading [28] and cooling under constant applied stress [27].

## 2. Experimental procedures

NiTi superelastic wire produced by Fort Wayne Metals in a cold-worked state (FWM #1 Ti-50.9 at. % Ni, 42% cold work, wire diameter 0.1 mm) was heat treated using a short pulse of electric current [5] (power density 160 W/mm³, pulse time 15 ms). During the heat treatment, a 30 mm long segment of cold-worked wire was crimped by two steel capillaries, prestressed to 400 MPa, constrained in length, and subjected to a 15 ms pulse of controlled electric power. The 15 ms heat-treatment gave rise to superelastic NiTi wire with fully recrystallized microstructure with a mean grain size of 250 nm (Fig. 1a), which displayed B2-R-B19' martensitic transformation with characteristic transformation temperatures of $M_s$ = -80 °C, $A_f$ = -27 °C, $R_s$ = -25 °C, and demonstrated superelasticity at room temperature (Fig. 1b). The stress-strain-temperature response of the 15 ms NiTi#1 wire in cyclic thermomechanical loads was highly unstable (Fig. 1c,d), as required for the planned experiments.

The isothermal (isostress) tensile tests (Fig. 2) were performed using a DMA 850 tester from TA Instruments with 10 mm long 15 ms NiTi #1 wire samples in position (strain) control. Special thermomechanical loading tests (Fig. 3) were conducted to evaluate the plastic strains generated separately by the forward and/or reverse martensitic transformation during cooling and/or heating under external tensile stress. The thermomechanical loading tests were performed in force and temperature control mode, with a cooling rate of 0.6 °C/s. Thermal strains were calibrated using a quartz sample and subtracted from the recorded signal.

## 3. Results

### 3.1. Stress-temperature diagram

The functional properties of NiTi, derived from cubic to monoclinic martensitic transformation, are well characterized by the stress-temperature (σ-T) diagram, which indicates the critical [temperature, stress] conditions for martensitic transformation and martensite reorientation processes. These conditions are evaluated from the isothermal stress-strain and/or isostress strain-temperature curves recorded in tensile tests on NiTi wires. Recently, we extended this standard approach by also including the critical conditions for plastic deformation of austenite and martensite into the σ-T diagram [33].

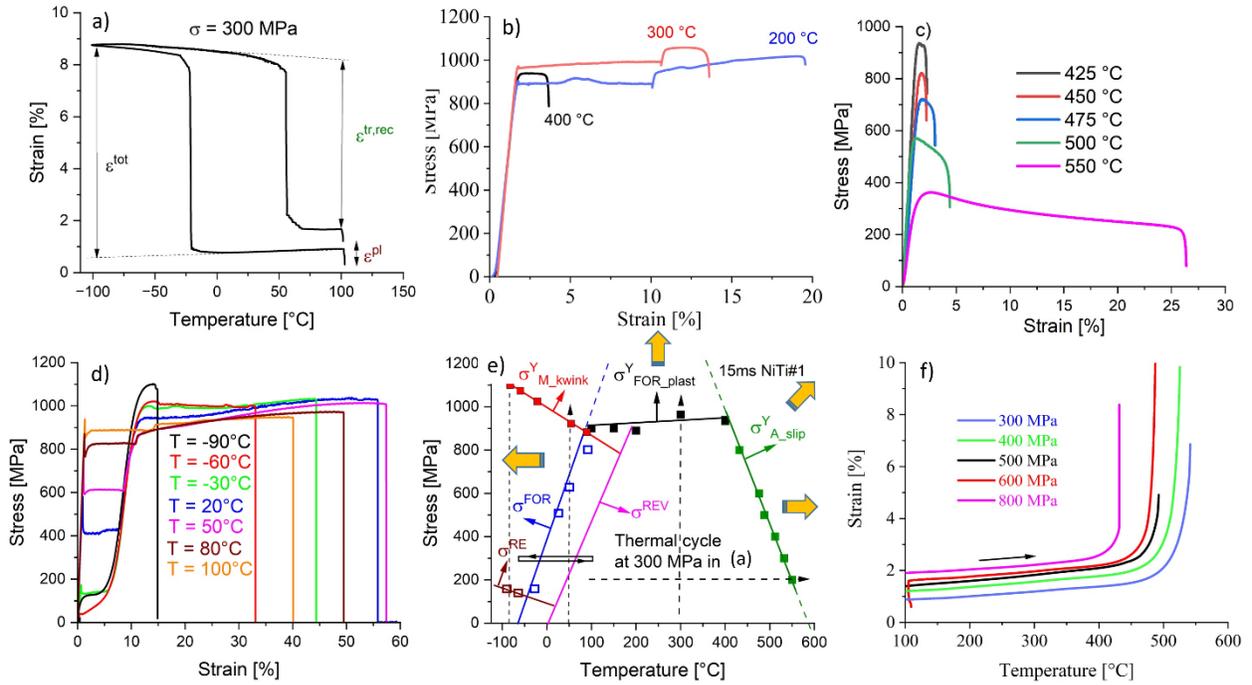

**Figure 2: Martensitic transformation and plastic deformation processes in superelastic 15ms NiTi #1 wire in tension**. a) strain-temperature response in thermal cycle under 300 MPa stress, b,c,d) stress-strain curves from isothermal tensile tests at various constant temperatures until fracture, f) strain-temperature curves from isostress heating under various constant stresses, e) stress-temperature diagram showing critical [temperature, stress] conditions for activation of various deformation/transformation processes in the wire determined from tensile tests (b,c,d,f).

Basic isothermal (Figs. 2b,c,d) and isostress (Figs. 2a,f) tensile tests were performed on the 15 ms NiTi #1 superelastic wire. Critical [temperature, stress] conditions for forward ($\sigma^{FOR}$) and reverse ($\sigma^{REV}$) martensitic transformations, for martensite reorientation ($\sigma^{RE}$) and for plastic deformation of austenite by dislocation slip ($\sigma^Y_{A\_slip}$) and plastic deformation of martensite by kwinking ($\sigma^Y_{M\_kwink}$) were evaluated. The isostress thermal heating under constant stress until fracture (Fig. 2f) was found to be a very useful experiment for determining the temperature-dependent yield stresses for plastic deformation of both austenite and martensite phases (see Fig. 4 in Ref [33]).

The σ-T diagram of the 15 ms NiTi #1 superelastic wire (Fig. 2e) shows these critical conditions. All deformation/transformation mechanisms, except of the plastic deformation of martensite by kwinking, are well known. Kwinking deformation was recently discovered as a mechanism of plastic deformation of B19' martensite in NiTi proceeding via combination of dislocation slip based kinking and twinning [36,37,38]. Surprisingly, little attention has been paid to plastic deformation of austenite by dislocation slip (Fig. 2c,f).

A special regime is observed in tensile tests at temperatures between 100°C and 400°C, where the wire deforms at ~900 MPa stress by martensitic transformation coupled with plastic deformation ($\sigma^Y_{FOR\_plast}$), even if neither the blue transformation line nor the green deformation lines are crossed in isothermal tensile tests (Fig. 2b). Tensile deformation tends to be localized in Lüders band fronts propagating in stress plateaus up to 400°C (Fig. 2b). However, ductility sharply decreases with increasing temperature, reaching only ~1% at 450°C (Fig. 2c). The deformation mechanism in this temperature range is not well understood. Notably, the theoretical critical stress for stress-induced MT ($\sigma^{FOR}$) (dashed extension of the forward transformation line in Fig. 2e according to the Clausius-Clapeyron equation) exceeds the critical stress to deform the induced martensite plastically ($\sigma^Y_{M\_kwink}$) above ~100°C (Fig. 2e). Although this seems contradictory, it can be viewed as stress-induced martensite deforming plastically immediately after it has been created from the austenite. We have clear experimental evidence that the stress induced MT occurs in tensile test at 150 °C [26] and indirect evidence that this is the case up to 400 °C (e.g. stress plateau in Fig 2b shall not exist without the MT). However, the stress induced MT is incomplete within the stress plateau range. The amount of stress-induced martensite sharply decreases with increasing temperature, leading to coupled martensitic transformation and plastic deformation in tensile tests at high temperatures when the $\sigma^Y_{FOR\_plast}$ line is crossed (Fig. 2e). The deformation mechanisms activated in these tests (Fig. 2b) are beyond the scope of this work and will not be discussed. For more information on this peculiar deformation mechanism, see Refs. [26,27].

The austenite deforms plastically at high temperatures above 400°C via dislocation slip in both isostress (Fig. 2c) and isothermal (Fig. 2f) tests. The stresses (temperatures) at which plastic deformation via dislocation slip in austenite occurs decrease with increasing temperature, as shown by the green line in the σ-T diagram in Fig. 2e. The stress-strain curves of the wire in this temperature range are characterized by: i) flow stress decreasing with increasing temperature, ii) strain softening, and iii) strain rate sensitivity (Fig. 2c). It is important to note that the critical conditions for plastic deformation of austenite ($\sigma^Y_{A\_slip}$) are significantly different from the stresses and temperatures at which martensitic transformations occur in cyclic thermomechanical loads on this NiTi wire (Fig. 2a). In fact, the 15 ms NiTi #1 superelastic wire does not deform plastically below 900 MPa in the 100-400°C temperature range, despite its fully recrystallized austenite microstructure. Based on this, we can reasonably conclude that this wire should not deform via dislocation slip in austenite when thermally cycled at 300 MPa stress (Fig. 2a).

The key point is that the NiTi wire should not deform plastically if the stress and temperature vary within the "generalized elasticity area" below the critical conditions for plastic deformation in the σ-T diagram (Fig. 2e) - only elastic deformation and deformation/transformation processes derived from martensitic transformation should be activated. However, this is not the case, the wire exhibits incremental plastic strains whenever the forward and reverse martensitic transformations proceed under external stress, even deep within this generalized elasticity area, as clearly evidenced in Figs. 1c,d, and 2a.

The incremental plastic deformation accompanying martensitic transformation under stress is known in the SMA field to be the origin of functional fatigue of NiTi [2,4,5,9,11]. For the purpose of this research, we consider it as a kind of coupled martensitic transformation and plastic deformation, which is activated whenever the forward and/or reverse MTs occurs under external stress. Plastic deformation via this deformation mode takes place whenever the external stress and temperature applied in a thermomechanical load cross one of the transformation lines $\sigma^{FOR}$ or $\sigma^{REV}$ in the σ-T diagram in the directions of arrows at the transformation lines (Fig. 2e).

**3.2. Recoverable and plastic strains generated by forward and reverse transformations under external stress**

It is logical to believe that plastic deformation occurs during the forward MT, particularly in isothermal tests on NiTi, where the upper plateau stress is significantly higher than the lower plateau stress due to the wide hysteresis of NiTi. On the other hand, plastic deformation would be expected to proceed more likely upon heating in the case of thermal cycling under constant stress, since dislocation slip is deemed to be easier at higher temperatures. These conflicting expectations create uncertainty about whether plastic strain generation occurs during the forward and/or reverse MT.

Since both forward and reverse MTs proceeding under stress in the standard thermal cycle (Fig. 3a,b) generate plastic strains, it becomes challenging to determine how much of the recorded unrecovered plastic strain ($\varepsilon^{pl} \approx 1.5\%$ in Fig. 3b) was generated during the forward MT and how much during the reverse MT. Additionally, the recoverable transformation strain evaluated from the standard test ($\varepsilon^{tr,rec} \approx 7\%$ in Fig. 3b) differs from the recoverable transformation strain generated by the forward MT $\varepsilon_F^{tr,rec}$, as will be explained below. However, when investigating the actuation performance of NiTi wire and evaluating the accumulation of residual strains upon thermal cycling, the $\varepsilon^{tr,rec}$ and $\varepsilon^{pl}$ strains are perfectly relevant. On the other hand, when we want to investigate the mechanism by which plastic strains are generated, the forward and reverse MTs have to be dealt with separately.

To evaluate the plastic strains generated by the forward and reverse MTs separately, we conducted thermomechanical loading experiments involving heating/cooling under stress according to the method (Fig. 3) originally introduced by Heller et al. [10] and described in detail in the Supplementary Materials (S1-S7). Two conditions need to be met for the method to be applied. First is that MTs proceeding in the absence of external stress (during cooling or heating under low applied stress of 20 MPa) do not generate plastic strains, and second is that tensile deformation of the wire in the low-temperature martensite state up to the yield stress must be fully recoverable upon stress-free heating (Fig. S8a,b).

The method involves running closed loop thermomechanical loading tests counterclockwise (Fig. 3c,d) and clockwise (Fig. 3e,f) along the rectangular path in the σ-T diagram to evaluate unrecovered plastic strains generated by the forward MT upon cooling under external stress (Fig. 3d) and during the reverse MT upon heating under external stress (Fig. 3f), respectively. The evaluated transformation strain $\varepsilon_F^{tr,rec}$ ($\varepsilon_R^{tr,rec}$) and plastic strains $\varepsilon_F^{pl}$ ($\varepsilon_R^{pl}$) generated by the forward MT (reverse MT) are associated with temperatures and stresses at which the forward and reverse MTs take place (Fig. 3ac,e,d,f). They are clearly different from the recoverable transformation strain $\varepsilon^{tr,rec}$, and plastic strain $\varepsilon^{pl}$ evaluated from the standard thermal cycle (Fig. 3a,b).

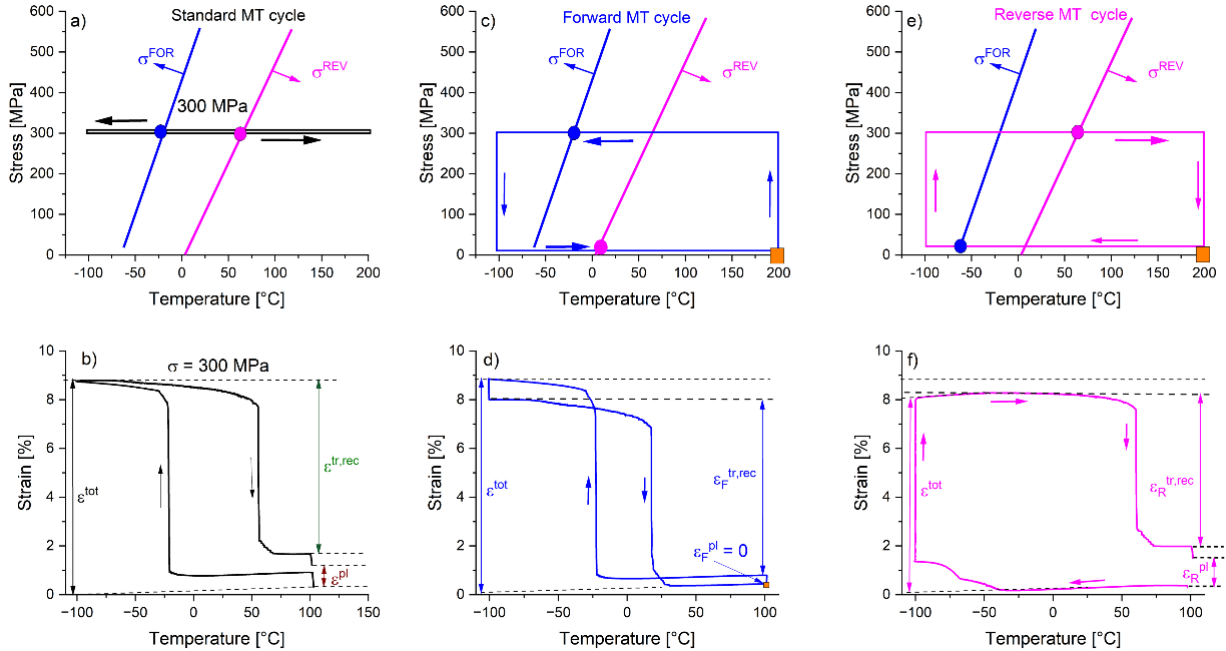

**Figure 3: Special thermomechanical loading tests to determine unrecovered plastic strains generated separately by the forward and reverse MTs in NiTi wire in tension [10].** Stress-temperature paths in σ-T diagram (a,c,e) and strain-temperature response of the wire (b,d,f) recorded in thermomechanical loading tests. Standard test involving cooling-heating under constant 300 MPa stress (a) is used to evaluate total strain $\varepsilon^{tot}$, transformation strain $\varepsilon^{tr,rec}$, and plastic strain $\varepsilon^{pl}$ generated in the thermal cycle (b). The tests along rectangle temperature-stress paths (c,e) are used to evaluate total strain $\varepsilon^{tot}$, transformation strain $\varepsilon_F^{tr,rec}$ ($\varepsilon_R^{tr,rec}$) and plastic strains $\varepsilon_F^{pl}$ ($\varepsilon_R^{pl}$) generated by the forward MT (reverse MT) as shown in (e,f).

Plastic strains generated by forward and reverse MTs in 15 ms NiTi #1 wire were experimentally evaluated for a wide range of external stresses 20-800 MPa (Fig. 4). It is worth noting that the strains due to MTs evolve at nearly constant temperatures. If not, temperatures corresponding to the mid of the transformation interval are taken as temperatures at which the MTs occurred. Another observation worth mentioning is the unusual increase of strain observed upon heating under low stresses 60-100 MPa in the Reverse MT tests (magenta curves in Figs. 4c,d,e). This unusual strain increase upon heating is attributed to thermally induced martensite reorientation [20]. When the self-accommodated martensite in the NiTi wire is heated under low stresses 60-100 MPa, it becomes unstable against the applied tensile stress, and when the stress-temperature path crosses the reorientation line in the σ-T diagram (Fig. 2d), martensite reorientation takes place and sudden strain increase is recorded (Fig. 5a). It shall be noted that the total strains induced by cooling under constant stress $\varepsilon^{tot}$ are systematically slightly larger than total strain $\varepsilon^{tot}$ reached by tensile deformation up to the same stress at -100 °C in martensite state.

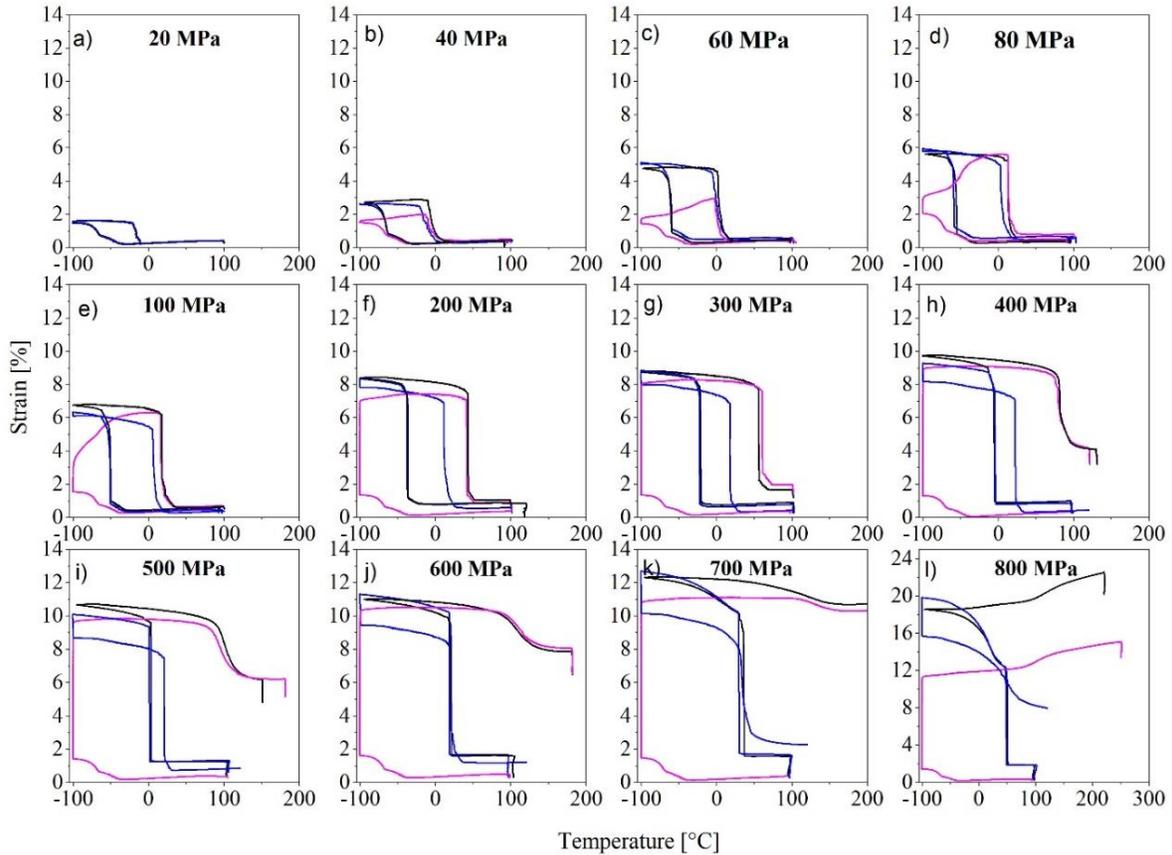

**Figure 4:** **Strain-temperature curves evaluated in set of thermomechanical loading tests performed at constant applied stresses 20-800 MPa according to the scheme introduced in Fig. 3.** The strain-temperature curves recorded in Standard cooling-heating under constant stress (black curve), Forward MT test (blue curve) and Reverse MT test (magenta curve) are plot together. Strain–temperature response of the wire in thermal cycle under 20 MPa was included for completeness.

The results of thermomechanical loading tests from Fig. 4 are presented in Fig. 5. The values of transformation temperatures, at which the forward and reverse martensitic transformations take place (blue and magenta solid circles in Fig. 5a) follow linear stress-temperature dependence with slopes s=6.9 MPa/K and 4.5 MPa/K, respectively. The [temperature, stress] data in Fig. 5a were taken from the Forward MT and Reverse MT tests, but they would be similar, if taken from the Standard tests (since the reverse transformation temperatures in Standard MT and Reverse MT tests are nearly the same (Fig. 4)).

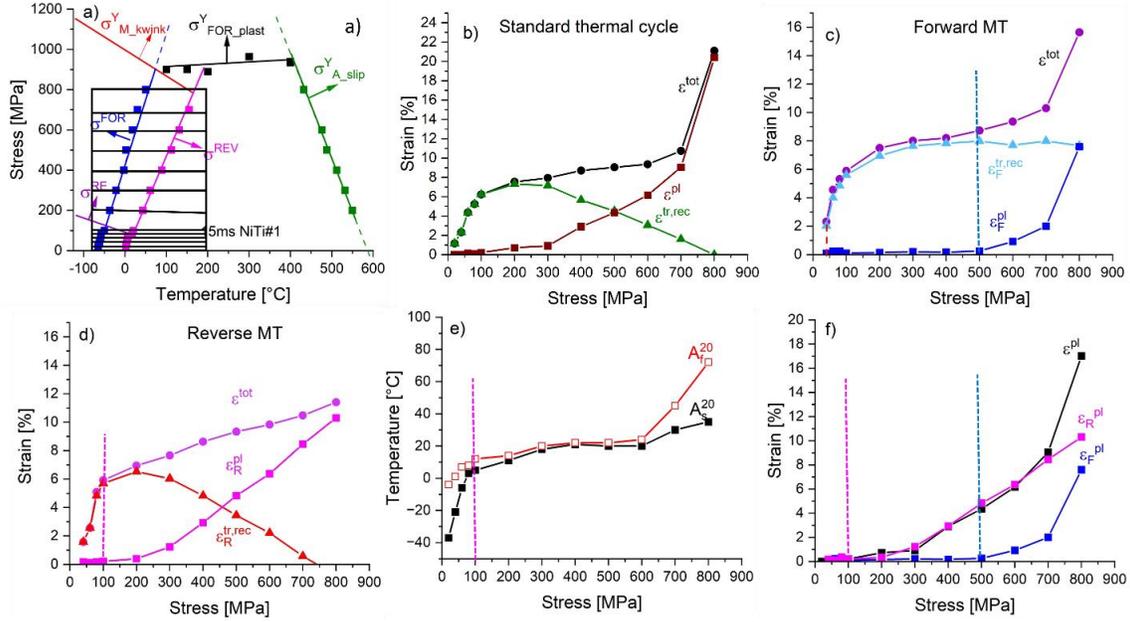

**Figure 5: Strains generated by forward and reverse MTs in 15 ms NiTi #1 superelastic wire thermally cycled under constant stresses 40-800 MPa (Fig. 4)**. a) Stress-temperature diagram with denoted [temperature, stress] conditions at which the forward (blue points) and reverse (magenta points) MTs take place. Figures (b,c,d) show stress dependence of total strain $\varepsilon^{tot}$, recoverable transformation strains $\varepsilon^{tr,rec}$, $\varepsilon_F^{tr,rec}$, $\varepsilon_R^{tr,rec}$ and plastic strains $\varepsilon^{pl}$, $\varepsilon_F^{pl}$, $\varepsilon_R^{pl}$ evaluated in Standard thermal test (b), Forward MT test (c) and Reverse MT tests (d), respectively. e) Reverse transformation temperature $A_S$ and $A_F$ evaluated upon heating under 20 MPa stress in Forward MT tests. f) Comparison of plastic strains generated by the forward and reverse MTs under various applied stresses. The blue and magenta vertical lines denote maximum stresses at which the forward and reverse MTs do not generate plastic strains.

The total strains $\varepsilon^{tot}$ generated by the forward MT upon cooling under applied stress (Fig. 5b,c) increase with increasing applied stress up to ~20% strain. It is important to note that the total strain $\varepsilon^{tot}$ is partially due to the oriented martensite [27] and partially due to plastic deformation generated by the forward MT upon cooling under stress, as will be discussed below. A different situation is in the Reverse MT test, in which the total strains $\varepsilon^{tot}$ (Fig. 5d) were not generated by the martensitic transformation but by martensite reorientation during the tensile deformation of the martensitic wire at -100 °C up to the desired tensile stress (stages 1 and 2 in Reverse MT cycle in Fig. 3) and/or during heating under low stress (Fig. 4).

Fig. 5e shows the reverse transformation temperature $A_s^{20}$ and $A_F^{20}$ evaluated upon heating the unloaded wire under 20 MPa stress in dependence on the stress applied on cooling (in Forward MT tests). Notably, these temperatures exhibit a sharp increase with increasing stress below 100 MPa and above 600 MPa, but only a slight increase in between. The rise in transformation temperatures at low stresses below 100 MPa is attributed to martensite stabilization by deformation (Figs. S4-S7). It is important to note that martensite stabilization by deformation below 100 MPa stress does not necessarily involve plastic strain via dislocation slip (Fig. S8a,b). The observation that the reverse transformation temperatures do not depend on the external stress applied on cooling in the stress range 100-600 MPa (Fig. 5e) is believed to be closely linked to the habit plane of the reverse MT proceeding in the absence of stress, as discussed in section 4.5. However, when the forward MT occurs at the highest stresses above 600 MPa, the reverse transformation temperatures increase due to the permanent lattice defects introduced by very large plastic strains generated by the forward MT upon cooling under high stresses (Fig. 5f).

The experiment demonstrates that plastic strains are generated by both forward MT during cooling and reverse MT during heating (Fig. 5f). While the forward MT begins to generate plastic strains $\varepsilon_F^{pl}$ at ~500 MPa stress (Fig. 5c), the reverse MT starts to generate plastic strains $\varepsilon_R^{pl}$ already at 100 MPa stress, at which martensite reorientation takes place (Fig. 5d). Additionally, the recorded plastic strains increase with increasing temperatures in different ways, as depicted in Fig. 5f.

The recoverable transformation strain generated by the forward MT $\varepsilon_F^{tr,rec}$ (Fig. 5c) increases with increasing stress, reaching a maximum strain of ~7% at ~300 MPa, and then remains approximately constant with further increasing stress. The recoverable transformation strain generated by the reverse MT $\varepsilon_R^{tr,rec}$ also increases with increasing stress but reaches the maximum strain ~7% already at ~200 MPa, before decreasing to zero at 800 MPa (Fig. 5d). Why different recoverable transformation strains are evaluated in the Forward MT and Reverse MT tests if there is only a single transformation strain $\varepsilon_F^{tr,rec}$ generated either during the forward MT on cooling or during the tensile loading in martensite state in the Reverse MT tests. The reason lies in that the recoverable transformation strains $\varepsilon_R^{tr,rec}$ evaluated in the Reverse MT tests are modified by the plastic strains $\varepsilon_R^{pl}$ generated by the reverse MT, as will be further discussed in section 4.7.

The key findings from the experiments in Figs. 2,4,5 are that the 15 ms NiTi #1 wire undergoes plastic deformation not only when the applied stresses and temperature exceed the critical [temperature, stress] conditions for plastic yielding, as denoted by the red, black, and green lines in the σ-T diagram (Fig. 5a), but also when the forward and reverse MTs proceed under stress above certain characteristic stress thresholds. Martensitic transformations occurring under specific stresses generate plastic strains of characteristic magnitudes (Fig. 6b,c). These thresholds and characteristic magnitude of plastic strains generated by the forward and reverse MTs (Fig. 6) represent the functional fatigue limits of the 15ms NiTi

#1 wire. In other words, stress-strain-temperature responses recorded in cyclic closed-loop thermomechanical load tests are stable if the forward and reverse MTs proceed below these thresholds. Conversely, they become unstable, displaying predictable ratcheting, if the MTs proceed above these thresholds. It's noteworthy that the functional fatigue limits (Fig. 6) are specific to each NiTi wire. For comparison, refer to the superelastic NiTi wire with smaller grain size and partially recrystallized microstructure (Fig. 9 in [10]).

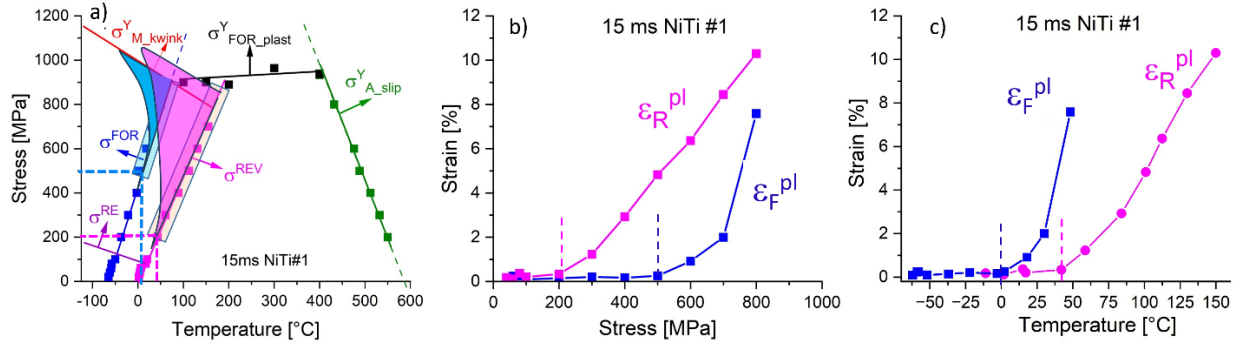

**Figure 6: Plastic strains generated by forward and reverse MTs in 15 ms NiTi #1 superelastic wire subjected to thermomechanical loading tests.** a) σ-T diagram with marked [temperature, stress] states in which the forward and reverse MTs occurred. The [temperature, stress] thresholds for generation of plastic strains are marked by dashed lines. The characteristic magnitudes of plastic strains generated by the forward and reverse MTs are plotted in dependence on the applied stress (b) and temperature (c) at which the MTs transformation occur.

## 4. Discussion

The research aimed at understanding the mechanism by which MT generates plastic strains seeks to inform strategies for modifying the austenitic microstructure of NiTi wires to suppress functional fatigue while maximizing reversible strains and expanding the range of stresses and temperatures applicable in cyclic thermomechanical loads. Conventional approaches to strengthening NiTi-based alloys against functional fatigue have typically focused on grain size refinement through processes like cold work/annealing [39,40], microstructure control via partial amorphization [40,41], $Ti_3Ni_4$ precipitation annealing [1,42,43], adjusting chemical composition to optimize lattice parameters for controlling strain-compatible habit planes [44], and designing tailored NiTi nanocomposites [45]. However, a challenge arises from the empirical nature of these strategies — the underlying deformation mechanism against which NiTi alloys are to be strengthened remains unknown. While the search for optimized lattice parameters to control strain-compatible habit planes [44] is theoretically sound, its relevance to the strengthening and functional fatigue of NiTi-based SMAs may be limited.

The success of strengthening NiTi to suppress functional fatigue is quantified by the experimentally observed stability of the cyclic stress-strain and/or strain-temperature responses in closed-loop thermomechanical loads, specifically through the accumulation of residual strains. A problem is the lack of understanding of deformation mechanism against which the alloy needs to be strengthened. This so far hindered the identification of a single material parameter which would quantify the propensity of NiTi alloy to functional fatigue. A promising candidate might be the yield stress for plastic deformation of martensite by kwinking (Fig. 2e), as it is clearly linked to cyclic instability [5]. However, the relationship between kwinking deformation, which activates at high stresses [36,37], and cyclic instability of thermomechanical responses at low stresses is not yet understood. Therefore, our objective is to first elucidate the mechanism by which the forward and reverse MTs proceeding under external stress generate plastic strains. Once this mechanism is understood, we can then explore potential strategies to strengthen NiTi alloys against this specific deformation mechanism.

Of course, from an application point of view, high-strength NiTi wires [43], which exhibit stable responses in cyclic thermomechanical loads for thousands of cycles, as e.g. 8-12 ms NiTi #1 wires (Figs. 9,13 in [39]), are of primary interest. However, we could not use these wires in our experiments because the plastic strains generated by a single martensitic transformation (MT) would be too small to be reliably evaluated. Instead, we used the superelastic 15 ms NiTi #1 and shape memory NiTi #5 wires, which display large plastic strains generated by single forward and/or reverse MT. The plastic strains generated by MTs under stress gradually decrease upon cycling. In this work, we simply assume that higher plastic strain generated in the first closed loop cycle correlates with higher cyclic instability (see e.g., Figs. 5,6 in Ref. [43]). Additionally, since the austenitic microstructure of the 15 ms NiTi #1 wires is free of dislocation defects (Fig. 1a), this wire was used for TEM analysis of dislocation defects created by single forward and reverse MTs [8].

The plastic deformation generated by the MT may occur in the austenite, martensite, or both phases. Plastic deformation of the B2 austenite is assumed to proceed via dislocation slip in the [100](011) or [1-11](011) slip systems, in accordance with the literature [46]. Such dislocations have been frequently observed in cyclically deformed superelastic NiTi [2,7,8,45]. Plastic deformation of the low-symmetry monoclinic martensite is less understood [34-38]. As far as we are aware, there is only the [100](001) dislocation slip system believed to be potentially activated in the martensite at low applied stresses [34,35,38] and kwinking deformation [36,37,38], which is responsible for plastic yielding of martensite at high applied stresses. Since the [100](001) slip dislocations are inherited by the austenite as [100](011) dislocations, it is challenging to determine whether they were created in the austenite or martensite.

Before discussing the plastic strains generated by the forward and reverse MTs, it is important to highlight a crucial experimental observation that was not previously mentioned. When the forward MT occurs at low stresses (20-500 MPa), it proceeds in a single stage at a constant temperature. However, at the highest stresses (600-800 MPa), the transformation upon cooling occurs in two stages, with strain further increasing upon cooling below the temperature plateau. Texture evolution studies during cooling under stress [27] reveal that this happens because the martensitic transformation is incomplete in the first stage and continues further in the second stage. This scenario is analogous to the forward MT localized in Lüders bands propagating during upper plateau stress in isothermal tensile tests at elevated temperatures [28]. Such forward MT was found to be incomplete within the plateau range and followed by further forward MT upon additional loading beyond the end of the stress plateau [28].

**4.1 Are the plastic strains generated by the forward and/or reverse martensitic transformation?**

It was shown in Section 3.2 that forward and reverse martensitic transformations (MTs) generate characteristic plastic strains above certain stress thresholds (Figs. 4-6). However, does this apply to general thermomechanical loads? Specifically, are the plastic strains generated by MTs under stress (Fig. 5f) also applicable to cyclic isothermal superelastic tests where forward and reverse MTs occur at the same temperature but different stresses?

To verify this for reverse MT, we performed thermomechanical loading experiments shown in Fig. 7. The wire was deformed up to the end of the stress plateau at 20 °C, as the forward MT at 450 MPa (Fig. 5c) does not generate plastic strains. The loading was stopped at the end of the stress plateau, and the wire was cooled (or heated) to a new temperature between -20 °C and 100 °C under constant strain. The wire was then unloaded at this new temperature (allowing reverse MTs to occur at different lower plateau stresses) and finally heated above the $A_f$ temperature to evaluate the plastic strains generated by the reverse MT.

Since the forward MT at 20 °C does not generate plastic strains, the plastic strains evaluated in the tests in Fig. 7a,b,c had to be generated only during the reverse MT on unloading. The lower plateau stress in tensile tests increases with increasing temperature (Fig. 7a,b,c) and gradually disappears so that the stress-strain response on unloading eventually becomes nearly linear. The magnitude of plastic strain generated by the reverse MT (Figs. 7a,b,c) is zero when unloading below 20 °C but increases with temperature, reaching up to 2% strain at 70 °C. However, there is a chance that plastic deformation might have occurred while the wire was heated under constant strain from 20 °C up to 70 °C (Fig. 7a). Independent experiments verified this, showing no plastic strain generated when the wire was heated to 70 °C and then cooled back to room temperature (Fig. S12). Therefore, the significant plastic strains observed in tensile tests with forward loading at 20 °C (Fig. 7a,b,c) were indeed generated solely by the reverse MT on unloading at 40-70 °C.

A similar experiment was conducted by deforming the wire at 60 °C. At this temperature, the forward MT at 650 MPa generates a plastic strain of 1.3% (Fig. 5c). Thus, the evaluated plastic strains (Figs. 7d,e,f) were generated by both forward and reverse MTs (Fig. 7f). The plastic strains recorded upon unloading at various temperatures are larger due to contributions from both MTs. These results demonstrate that the residual plastic strains observed during superelastic cycling of NiTi (Fig. 1c) are primarily generated by the reverse MT on unloading. This finding contradicts the common belief widely spread in the SMA field that the incremental plastic strains accumulating during the superelastic cycling (Fig. 1c) are mainly generated during the forward MT.

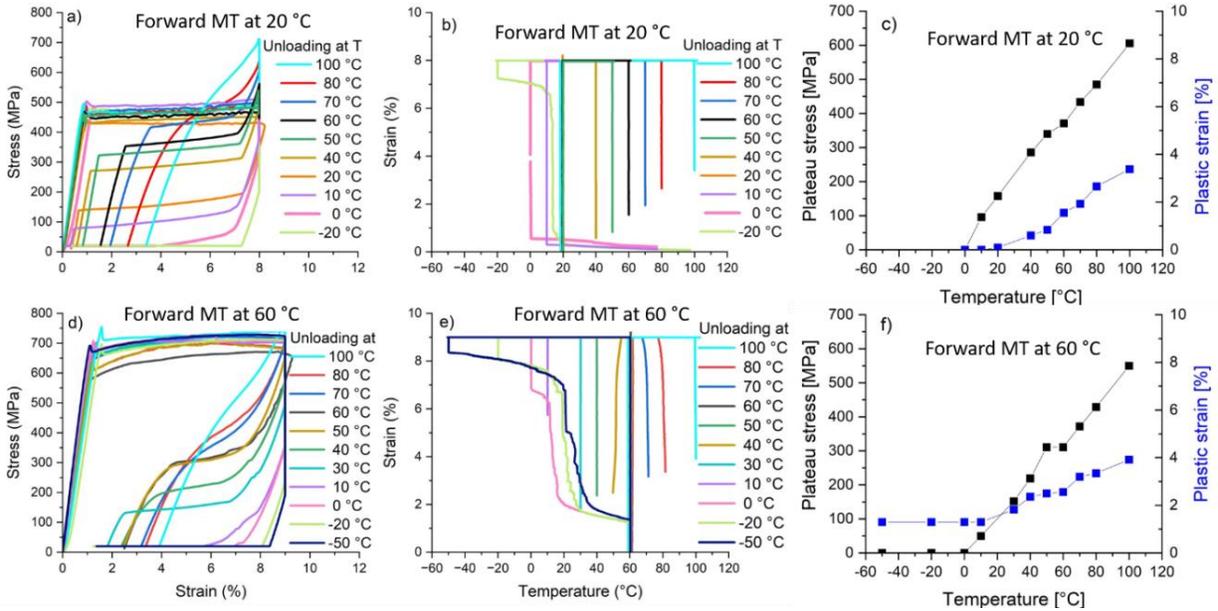

**Figure 7: Two sets of thermomechanical loading experiments designed to evaluate plastic strains generated by reverse MT on unloading in superelastic cycles on 15ms NiTi #1 wire**. The wire was deformed up to the end of the stress plateau at 20 °C (a,b,c) and 60 °C (d,e,f), the test was stopped, temperature was changed while strain was kept constant to a temperature within the range -20 °C – 100 °C, the wire was unloaded at this temperature wire and finally, the wire was heated above $A_f$ temperature up to 60 °C. Temperature dependence of lower plateau stress and plastic strain generated by the reverse MT on unloading and heating are shown in (c,f). In the first set (a-c) (forward MT occurred at Plastic strain 20 °C), no plastic strain was generated by the forward MT. In the second set (d-f), forward MT generated 1.3% strain.

### 4.2 Generation of plastic strains in cyclic thermomechanical loading tests

The presented experimental results (Figs. 4-6) suggest that there are certain stress thresholds, above which the forward MT and reverse MT start to generate plastic strains and that the magnitudes of such plastic strains increasing with increasing stress are well-defined (Fig. 6). Therefore, when designing engineering applications with NiTi wire, it is crucial to consider the generation of plastic strains separately for the forward and reverse MTs. In thermomechanical loading cycles, plastic strains can be avoided if both forward and reverse MTs occur below the stress thresholds (Fig. 6a), ensuring no functional fatigue. Exceeding these thresholds leads to strain accumulation proportionate to the characteristic strain magnitudes for each MT at the given stress (Fig. 6b,c). It is important to note that this oversimplifies the cyclic deformation process, as it ignores the gradual decrease in incremental plastic strain with each cycle (Fig. 1d).

To validate the proposed approach, we conducted cyclic loading experiments, some of which are presented in Fig. 8. The results affirm the idea that the cyclic stress-strain-temperature curves recorded in closed-loop thermomechanical tests stabilize when the external stress approaches the thresholds for generating plastic strains. In particular, the stability of the stress-strain response improves significantly with decreasing test temperature (Figs. 8b,c,d) in cyclic superelastic tests at 20, 10, and 3 °C (path 1 in Fig. 8a) and cyclic stability in thermal cycling under 200 and 100 MPa stress (path 2 in Fig. 8a) exhibits similar improvements (Figs. 8e, f). Despite applying 600 MPa stress and 10% strain in the shape memory tests, the stress-strain-temperature response remains stable because martensitic transformation does not occur under stress in this test.

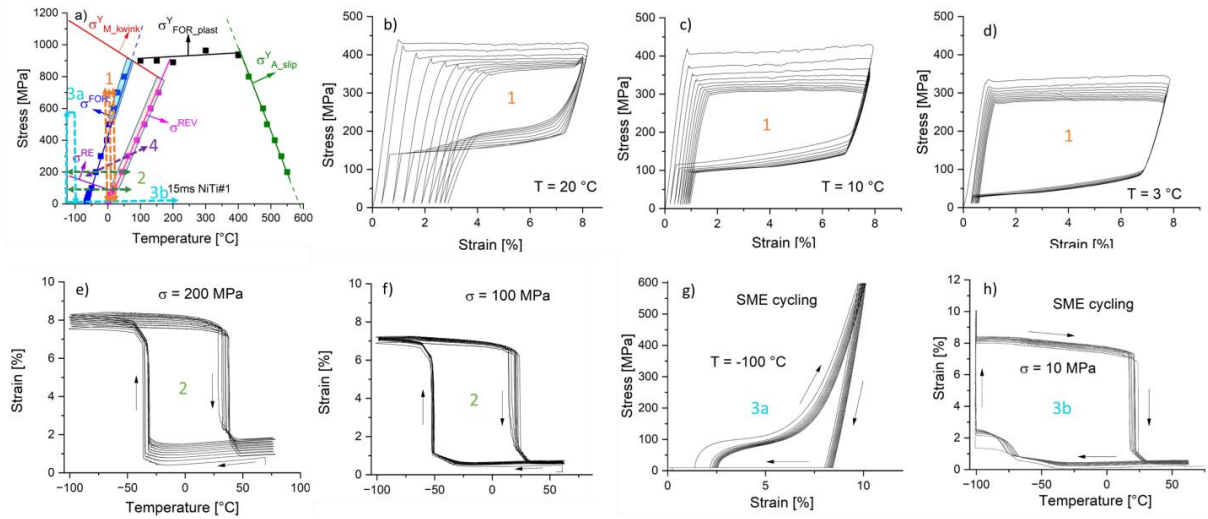

**Figure 8: Cyclic thermomechanical loading tests on 15ms NiTi #1 wire (10 cycles) near the stress threshold for generation of plastic strain by the forward and reverse MTs.** a) stress-temperature paths of closed loop thermomechanical loading cycles 1-4 in the σ-T diagram. Stress-strain responses in isothermal superelastic cycling at 20 °C (b), 10 °C(c) and 3 °C (d). Strain-temperature responses in isostress thermal cycling under 200 (e) and 100 MPa (f). Stress-strain-temperature response in shape memory cyclic test involving tensile loading-unloading up to 600 MPa (g) followed by heating-cooling up to 70 °C (h).

Considering that the 15 ms NiTi #1 wire was deliberately heat-treated to induce large plastic strains and significant cyclic instability (Fig. 1d), the remarkable cyclic stability observed in superelastic tests at 3 °C, thermal cycling under 100 MPa (Fig. 8f), and shape memory tests (Fig. 8g, h) is noteworthy. In addition, the proposed concept of functional fatigue limits (Fig. 6) sheds light on the challenge of achieving cyclically stable stress-strain-temperature responses in actuator tests (path 4 in Fig. 8a), where reverse martensitic transformation under stress is unavoidable.

In summary, our findings demonstrate that both forward and reverse MTs induce plastic strains of characteristic magnitudes only when surpassing specific stress thresholds (Fig. 6). To attain a cyclically stable stress-strain-temperature response from NiTi wire, we must either elevate the stress thresholds for plastic strain generation by strengthening the alloy or ensure that both forward and reverse martensitic transformations occur below these thresholds. The proposed concept of functional fatigue limits (Fig. 6) can serve as guidelines in engineering design to mitigate cyclic instability of NiTi wire under cyclic thermomechanical loads.

**4.3 How the plastic strains are generated by martensitic transformations under external stress**

We recognize that while we have observed plastic strains generated by both forward and reverse martensitic transformations, the exact mechanism underlying this phenomenon remains unclear. We would like to discuss this issue, but it is not possible based on the results of thermomechanical testing only. To shed light on this issue, we draw insights from our related studies focusing on martensite variant microstructures [27,28], martensite textures [26, 27], and lattice defects in austenite [8] created by forward and reverse martensitic transformations under stress in experiments involving superelastic and shape memory wires. Based on the findings presented in Fig. 2, we rule out the possibility of plastic deformation occurring in austenite prior to the forward MT, but not during the reverse MT under stress.

Given that martensite variant microstructures in the superelastic NiTi#1 wire cannot be analyzed by TEM at room temperature due to the martensite phase's instability at 20 °C, we reconstructed martensite variant microstructures created by forward martensitic transformation on cooling in grains of the 15 ms NiTi #5 SME wire, which is stable at room temperature [27]. Additionally, we determined the evolution of textures in austenite and martensite upon cooling the austenitic wire under various external stresses and observed lattice defects in martensite (Fig. 9). The austenite and martensite phases exhibit radially symmetric fiber textures (Fig. S9) effectively characterized by inverse pole figures (AD IPF), which depict the orientation distribution of the wire axis direction in cubic austenite and monoclinic martensite lattices.

In the absence of stress, the martensite variant microstructures observed in grains of the 15 ms NiTi #5 SME wire displayed multidomain (001) compound twin laminates and four-pole AD IPF textures (0 MPa in Fig. 9). However, with increasing applied stress, both the microstructure and martensite texture underwent changes, indicating a transition from self-accommodated martensite forming during stress-free cooling to oriented martensite forming via forward martensitic transformation under stress. At 100-200 MPa stresses (200 MPa in Fig. 9), the wire exhibited mainly single (001) compound twin laminates filling entire grains, with some detwinned grains and others containing (100) twins. The wire displayed a two-pole texture of oriented martensite reflecting the presence of (001) compound twinned grains [27]. Under higher stresses (400-600 MPa in Fig. 9), the wire showed a similar microstructure with additional (100) twin bands, along with a peculiar contrast attributed to the high density of slip dislocations. The martensite textures reflect the gradual reorientation of martensite between with applied stress increasing up to 200 MPa (increase of the intensity of the (10-3) pole) as well as the presence of (001) compound twins in the microstructure observed by TEM (tail of the (10-3) pole towards the [102] direction). Practically no change of the martensite texture occurs between 200 and 400 MPa evidencing no further change of the orientation of martensite variants with increasing external stress in this range. However, there is also the plastic deformation by dislocation slip that is seen neither by the TEM nor by the texture studies. The forward MT upon cooling the 15 ms NiTi #5 wire under 400 MPa generated 3% plastic strain (Fig. 9). In a contrast, the superelastic 15 ms NiTi #1 wire resisted plastic deformation up to 500 MPa stress (Fig. 5c).

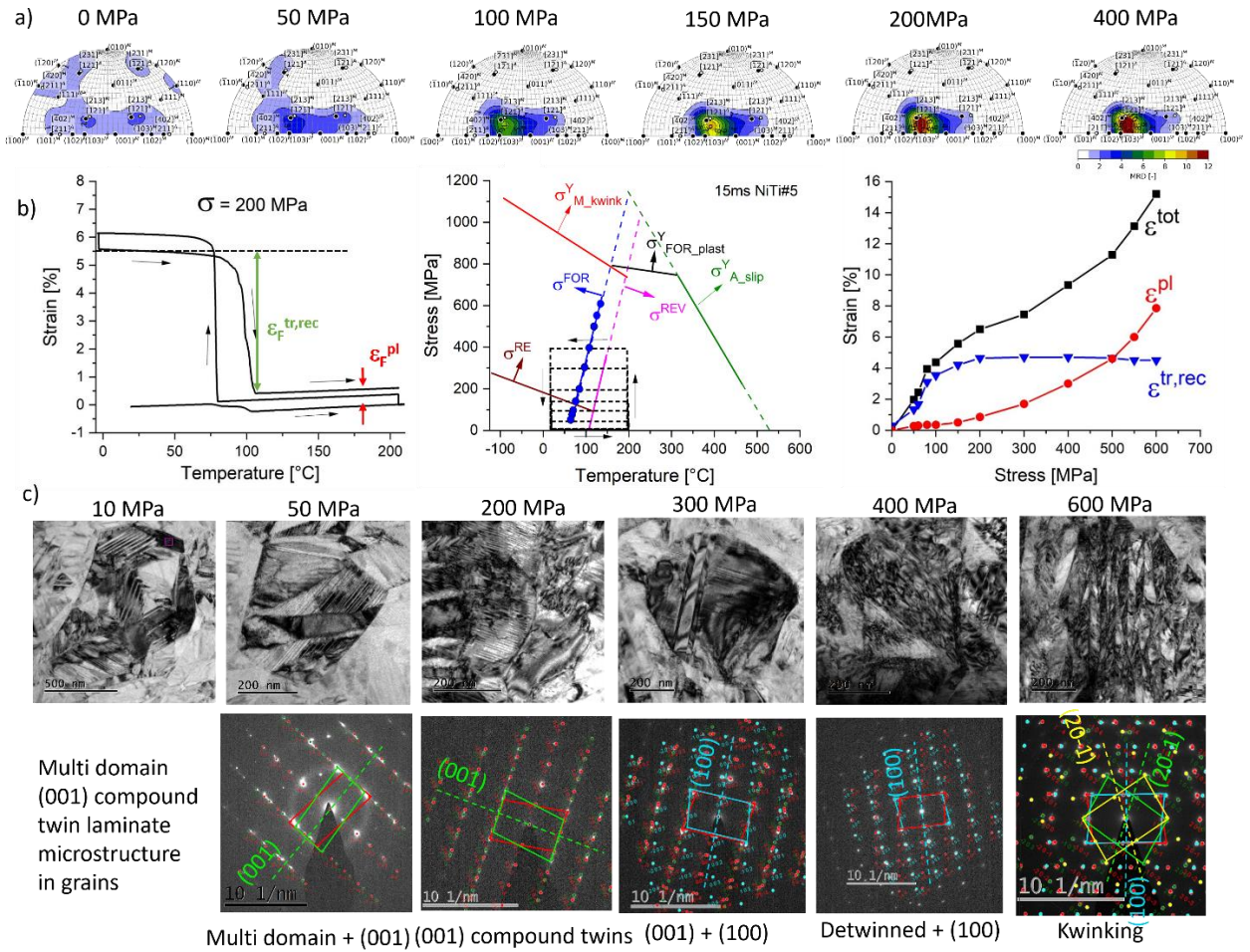

**Figure 9: Martensite textures (a), recoverable and plastic strains (b) and martensite variant microstructures (c) generated by single forward martensitic transformation upon cooling NiTi #5 SME wire under various external stresses [27].** Texture in martensite is characterized by AD IPF of the wire axis direction, recoverable and plastic strains were evaluated by closed loop thermomechanical cycles along rectangle paths in the σ-T diagram and martensite variant microstructures (BF image and composite electron diffraction pattern) in whole grains were evaluated by SAED-DF method in TEM. See Ref. [27] for details, particularly for reconstructions of martensite variant microstructures in grains of NiTi wires cooled under various external stresses.

Plastic deformation of martensite by kwinking, which introduced kwink bands into microstructure and brings about changes in martensite lattice orientations, on the other hand, can be easily detected by TEM as well as by texture studies. The martensite variant microstructures created by forward MT upon cooling under 500-600 MPa [27] contain high density of (20-1) kwink bands (600 MPa in Fig. 9) characteristic for plastic deformation of martensite via kwinking, with the forward MT generating 5% transformation strain and 5% plastic strain. Unfortunately, we lack texture data for martensite formed under stress exceeding 400 MPa, as these experiments were not conducted. Yet, further changes of martensite texture are anticipated upon cooling under 500-600 MPa based on the findings of texture evolution in austenite and martensite during tensile tests on 15 ms NiTi #1 wire at various temperatures, as explored in Ref. [26].

As a part of this research. we have investigated lattice defects left in the austenite phase induced by the forward and reverse MTs taking place under applied stresses in closed loop thermomechanical load cycles (Fig. 4). The results in a form of detailed TEM and HRTEM analysis of slip dislocations and deformation bands are reported separately [8]. The obtained results (summarized in Fig. 10) revealed several key points: i) lattice defects, predominantly slip dislocations at low stresses, were observed only if the MT occurs above the stress thresholds; ii) distinct types of slip dislocations emerge from the forward and reverse MTs; and iii) the density of lattice defects increases with increasing applied stress. Notably, while the forward MTs did not yield lattice defects up to 500 MPa (Fig. 10b), the reverse MT induced dislocation defects even at stresses as low as 200 MPa (Fig. 10e). Deformation bands, some of them being {114} austenite twins were observed in microstructures of wires cooled under 800 MPa and heated under 700 and 800 MPa.

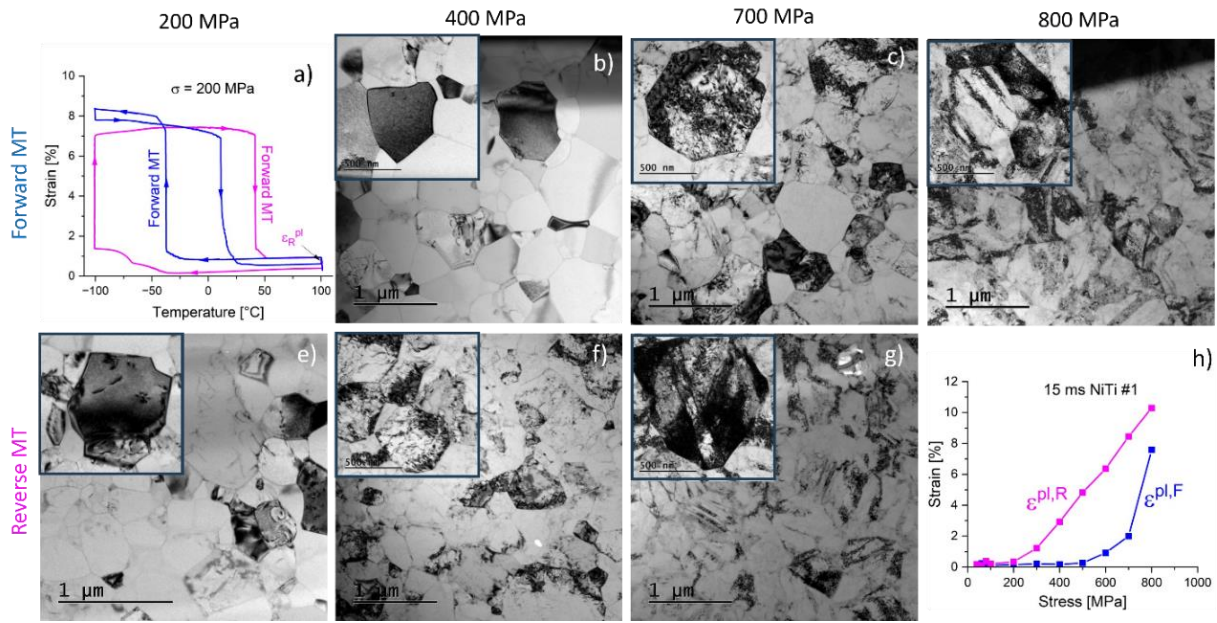

**Figure 10: Dislocation defects generated by single forward MT on cooling (b-d) and single reverse MTs on heating (e-g) of superelastic NiTi #1 wire under various applied stresses** [8]. Strain-temperature responses recorded in thermomechanical loading tests used to evaluate plastic strains generated by a single forward and reverse MT (a) and magnitudes of plastic strains generated by single forward and reverse MTs under applied stresses 0-800 MPa (h) were included for convenience of the readers. Bright field TEM images of dislocations created by forward and reverse MTs under 200.400.700 and 800 MPa stress. For detailed analysis of the dislocation defects see Ref. [8].

The slip dislocations generated by the forward and reverse MTs differ not only in density (higher density after reverse MT) but also in character. After forward MT, single dislocations are observed, whereas dense bundles of dislocation loops are generated by reverse MT. Austenite twin bands in the {114} planes were observed after forward MT at high stress 800 MPa (Fig. 10d) and after reverse MT under 700 MPa. Since these austenite bands originate from the plastic deformation of martensite by kwinking [22, 36, 37], we infer that the martensite undergoes plastic deformation via kwinking during both forward and reverse MTs only under the highest external stresses. This is consistent with the observation of very large plastic strains generated by both forward and reverse MTs proceeding under highest stresses (Fig. 10h).

In summary, the results of closely related experiments analyzing martensite variant microstructures, textures [27], and lattice defects in austenite [8] in deformed NiTi wires indicate that the oriented martensite created by forward MT under stress forms single (001) compound twin laminates filling entire grains. Such oriented martensite deforms plastically during the forward but mainly during the reverse MTs under stress. Consequently, we conclude that the plastic strains observed during thermal cycling of the superelastic NiTi #1 wire under constant applied stresses (Figs. 4,5) were primarily generated during the reverse MT on heating.

## 4.4 Why plastic strains are generated by martensitic transformations proceeding under external stress?

The 15 ms NiTi #1 wire does not deform plastically under stress up to ~900 MPa stress in tensile tests in a wide temperature range -100-400 °C (Fig. 2e). However, the reverse MT on heating under 300 MPa taking place at 65 °C generates 1.6% plastic strains (Fig. 3f, 6). There must be reason for the plastic deformation generated by the MT proceeding under such low applied stress. At the same time, MTs proceeding in the absence of stress (or under very low external stress) do not generate plastic strains regardless of previous history. Let's briefly explore why plastic deformation only occurs when the forward and/or reverse MT proceed under external stress.

The forward and reverse MTs proceed in tensile tests on NiTi wires in a localized manner via motion of cone shaped Luders band fronts (Fig. 11). It was reported in [48] that, local shear stresses within transforming grains on the Luders band front are about 20% higher than the shear stresses elsewhere within the untransformed austenitic part of the wire (Fig. 11). At the same time, since the cross section of the wire sharply decreases in locations where the Luders band front passes through, the stress induced martensite within the Luders band has to withstand significantly higher stress, otherwise the Luders band front would not propagate. As a result, plastic deformation occurs together with martensitic transformation within the propagating Luders band fronts. The macroscopic strain localization within the propagating Luders band fronts thus represents a strong argument explaining why plastic deformation occurs only when and where martensitic transformation proceeds in the tensioned NiTi wire.

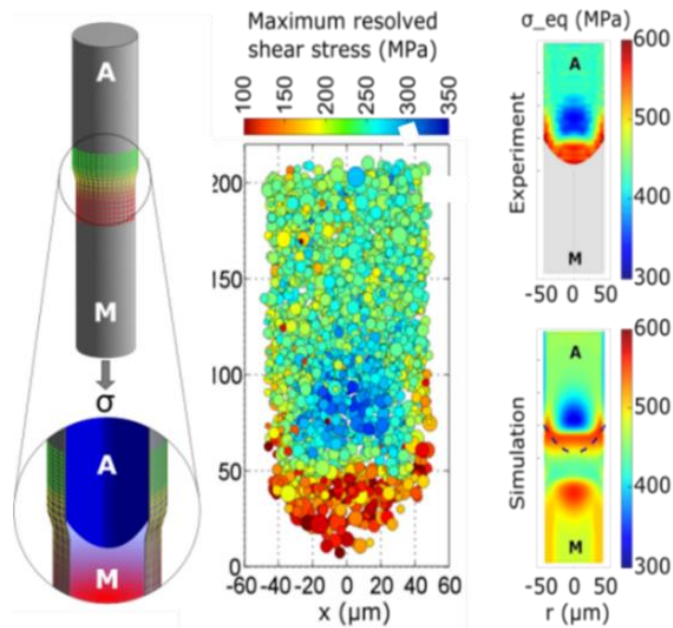

**Figure 11: Localization of deformation/transformation during tensile loading of NiTi wire on the cone shaped transformation front.** Up to 20% higher shear stress were evaluated on the transformation front surface by 3D-XRD experiment on superelastic NiTi #1 wire deformed in tension [48] compared to the shear stresses in other grains deformed only elastically.

However, this does not mean that the strain localization is the only reason. We believe it merely amplifies the coupling between MT and plastic deformation in NiTi. Even if martensitic transformation isn't macroscopically localized, MTs under stress generate plastic strains. We consider four potential reasons for this. The first reason is rooted in the fact that the martensite is plastically softer than austenite and may hence undergo plastic deformation once stress-induced from the austenite, contingent upon martensite's resistance against the [100](001) dislocation slip. This suggests that plastic deformation accompanies forward stress-induced MT, as particularly evident at stresses exceeding 500 MPa in the 15ms NiTi#1 wire, and it is highly heterogeneous at the mesoscale level of polycrystal grains (see section 4.6). The second reason arises from the plastic deformation of martensite aiding in achieving strain compatibility at stationary grain boundaries during forward MT under stress. The third reason involves plastic deformation of austenite or martensite assisting in strain compatibility at habit plane interfaces propagating during reverse MT (see section 4.5). Lastly, the fourth reason relates to oriented martensite heated under external stress being unable to withstand local stresses generated by reverse MT upon heating, leading to plastic deformation before or during the reverse MT [17,38]. Consequently, NiTi wire heated under high external stress elongates instead of shortening (Fig. 4l), as discussed in section 4.6.

## 4.5 Habit plane of the forward and reverse transformations proceeding under external stress

Based on observations of (001) compound twin laminates forming the martensite variant microstructures in grains of thermally induced self-accommodated martensite (Fig. 9), we assume that the forward as well as reverse MT upon stress-free cooling of the 15ms NiTi #1 wire proceeds via a habit plane between austenite and second-order (001) compound twin laminate of martensite, as proposed by Waitz [25]. The self-accommodated martensite is characterized by a 4-pole AD IPF texture (Fig. S9) [47]. Neither plastic strains (Figs. 4,5) nor lattice defects (Fig. 9) are generated by the forward and reverse MTs in the absence of external stress. The habit plane between austenite and second-order (001) compound twin laminate of martensite during the stress-free MT is considered regardless of the thermomechanical loading history – i.e., regardless of whether the martensite is self-accommodated, oriented, or plastically deformed [5,8]. It needs to be mentioned that we do not know crystallographic details of these habit plane interfaces. In contrast to nanograin microstructures (d ~ 20 nm) analyzed by Waitz [25], multiple variously oriented (001) compound twin domains appear in ~250 nm grains of the 15 ms NiTi #5 wire (10 MPa in Fig. 9) suggesting that the stress-free MT occurred via propagation of multiple habit planes in a single grain.

When the wire is cooled or heated under external stress (Figs. 5, 9, 10), however, plastic strains (Figs. 5,7) and lattice defects (Fig. 10e,f) become gradually generated by both forward and reverse MTs, though at different thresholds (Fig. 6). Martensitic transformation taking place under external stress must proceed via strain-compatible habit planes between austenite and martensite. As the strain compatible habit planes could differ for the forward MT and reverse MT and may also change with increasing stress, we present our view on the habit plane regimes at low (Fig. 12), medium (Fig. 13), and high (Fig. 14) stresses separately for the forward and reverse MTs (the stresses in Figs. 2-9 relate to 15 ms NiTi #1 wire).

When the NiTi wire is cooled under low stress 20-100 MPa (Fig. 12), the forward MT on cooling under stress proceeds via a habit plane between austenite and second-order (001) compound twin laminate of martensite, similar to stress-free cooling. However, there is a significant difference - the wire elongates during the forward MT upon cooling under stress and the martensite variant microstructures and textures become different from those observed during stress-free cooling (Fig. 9). As a result of this, suitable habit planes variants form preferentially and promptly reorient into single laminate of (001) compound twins (Fig. 12). No plastic strains and no lattice defects are generated by such forward MT under stress (Fig. 10).

Despite the large total strain generated upon cooling under 100 MPa stress (Figs. 4,5), no plastic strains (Fig. 5f) and no lattice defects (Fig. 10) were generated by the reverse MT under stresses of 0-100 MPa. Since the single (001) compound twin laminate forming the martensite variant microstructure in some grains cannot form a strain-compatible habit plane interface with the parent austenite, this martensite has to first reorient back into the second-order laminate (new unsuitably oriented (001) compound twinned martensite variants are introduced into the microstructure against the action of external stress) before a strain-compatible habit plane interface can form and transform the martensite back to the parent austenite. Based on the observed lack of plastic strains, we deduce that martensite reorients upon heating under low applied stresses below 100 MPa and the reverse MT proceeds via the same habit plane as the forward MT (Fig. 12). No slip dislocations are created during the reverse MT. However, since the external tensile stress acts against the martensite reorientation, reverse MT requires extra energy to be supplied via overheating. This is evidenced experimentally as the upward shift of the $A_s$ and $A_f$ temperatures with increasing stress (Fig. 5e) and represents the physical origin of the stabilization of martensite by deformation [20].

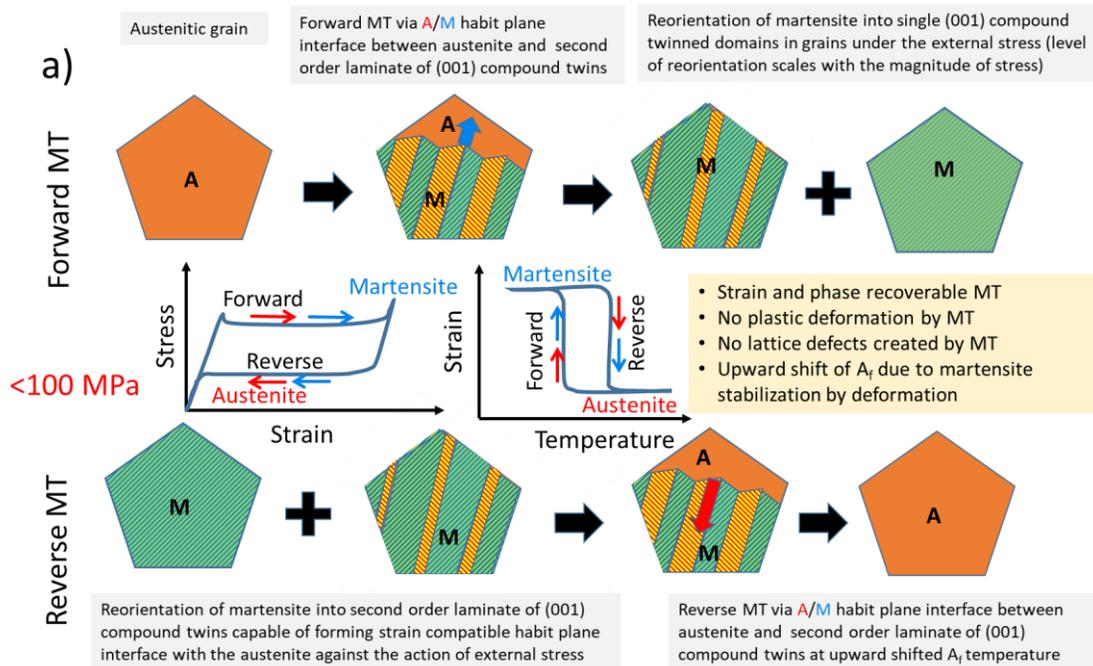

**Figure 12: Forward and reverse MTs under very low stress 10-100MPa.** Martensitic transformation proceeds via the same habit as during the stress-free cooling/heating across the transformation range. With increasing stress, martensite reorients more into a single (001) compound twin laminate. The reverse MT proceeds in an inverse manner without generating any dislocation defects since the strain compatible habit plane configuration can be achieved against the action of applied stress. $A_F$ temperature shifts upwards with increasing applied stress due to martensite stabilization by deformation.

When the NiTi wire is cooled under medium stresses ranging from 200 to 500 MPa (Fig. 13), the forward MT proceeds again via a habit plane between austenite and second-order (001) compound twin laminate of martensite. This induced martensite promptly reorients into a single (001) compound twin laminate (Fig. 9) without generating plastic strains (Fig. 5c) or lattice defects (Fig. 10b). This view is supported by TEM observations showing that most grains contain martensite variant microstructures consisting of single (001) compound twin laminates, with some grains detwinned and others containing (100) twins as well as martensite textures corresponding to oriented martensite (Fig. 9) [27].

In contrast, the reverse MT on heating under external stresses exceeding 200 MPa proceeds differently - it generates plastic strains (Fig. 5d) and lattice defects (Fig. 10) . The reason for this will be discussed in the next section 4.6. Simultaneously, the formation of a habit plane interface becomes challenging since the heated martensite is subjected to external stress and cannot reorient to form the required second-order (001) compound twin laminate for strain-compatible habit planes with austenite. Consequently, strain compatibility at the habit plane must be achieved differently. We assume that the reverse MT proceeds via a habit plane interface between a single variant of martensite (Fig. 13), assisted by elastic deformation as proposed by Heller and Sittner [31], and potentially with the aid of plastic deformation by dislocation slip in austenite [17].

Nevertheless, we do not think that the requirement for strain compatibility at the habit plane of the reverse MT is the key reason for the experimentally observed plastic strains generated by the reverse MT (Fig. 5d). Since plastic strains (Figs. 5d,7b) and slip dislocations (Fig. 10e) are generated by the reverse MT proceeding at stresses as low as ~200 MPa, it is likely that the reverse MT generates plastic strains when the external stress exceeds the reorientation stress 100 MPa (Fig. 2e). While this could be related to the habit plane of the reverse MT [49], it's not necessarily the sole factor; other reasons may also contribute, as discussed in section 4.4.

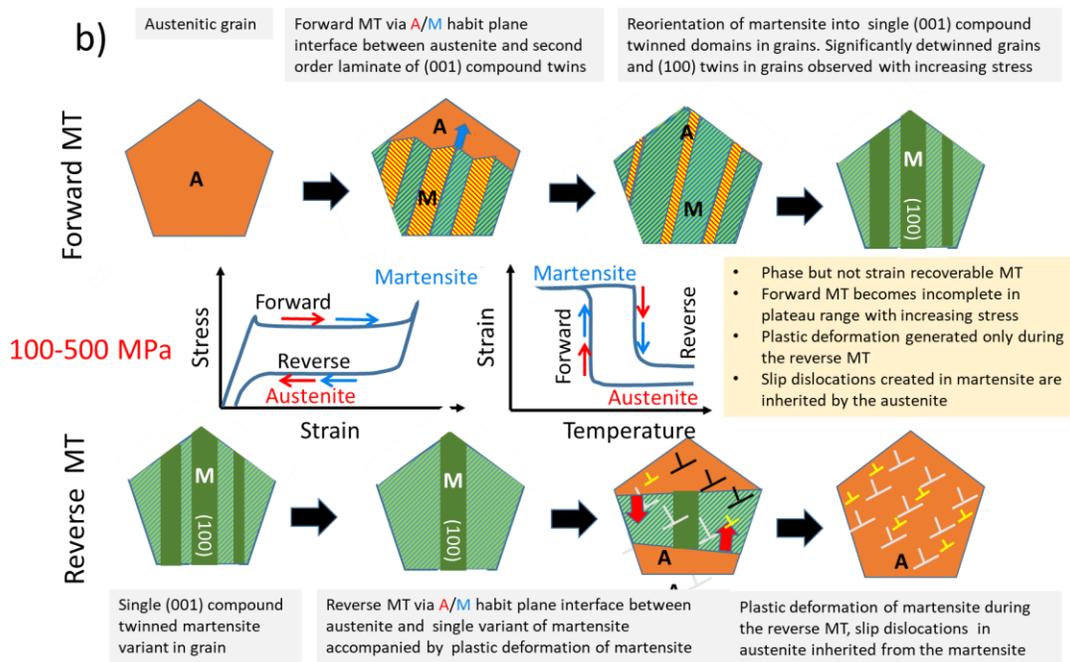

**Figure 13: Forward and reverse MTs under medium stress 200-500MPa.** The forward martensitic transformation proceeds via same habit as at low stresses, some grains detwin and (100) martensite twins appear in the microstructure. The reverse martensitic transformation proceeds via a habit plane between austenite and detwinned martensite with the assistance of elastic [31] and plastic deformation in austenite (yellow dislocations). In addition, plastic deformation in martensite occurs during the reverse MT, the magnitude of which increases with increasing applied stress (black dislocations in martensite are inherited into austenite as white dislocations).

When the NiTi wire is cooled under very high stresses (~800 MPa), we still presume that the forward MT proceeds via the habit plane between austenite and the second-order (001) compound twin laminate of martensite (Fig. 14). Main reason is the frequent observation of (001) compound twins in reconstructed martensite variant microstructures (Fig. 9). The induced martensite immediately reorients and deforms plastically through kwinking, as indicated by the presence of kwink bands within the microstructure (Fig. 9), the texture of stress-induced martensite [27,28], and the presence of {114} austenite twins in the wire's microstructure (Fig. 10d). The forward MT is not completed within the temperature plateau upon cooling under stress. While the retained austenite continues to transform upon further cooling under stress, the martensite undergoes plastic deformation through dislocation slip and kwinking to accommodate the strain incompatibilities at grain boundaries [28]. The activation of the kwinking deformation of martensite leads to very large plastic strains generated by the forward MTs (Figs. 4l, 5f).

The reverse MT upon heating under large stresses (~800 MPa) is assumed to proceed via a habit plane interface between austenite and detwinned martensite, assisted by elastic deformation [31] and plastic deformation via dislocation slip in austenite [17]. When plastic strains generated by the reverse MT exceed the recoverable transformation strain generated by the forward MT, the wire elongates during the reverse MT (Fig. 4l).

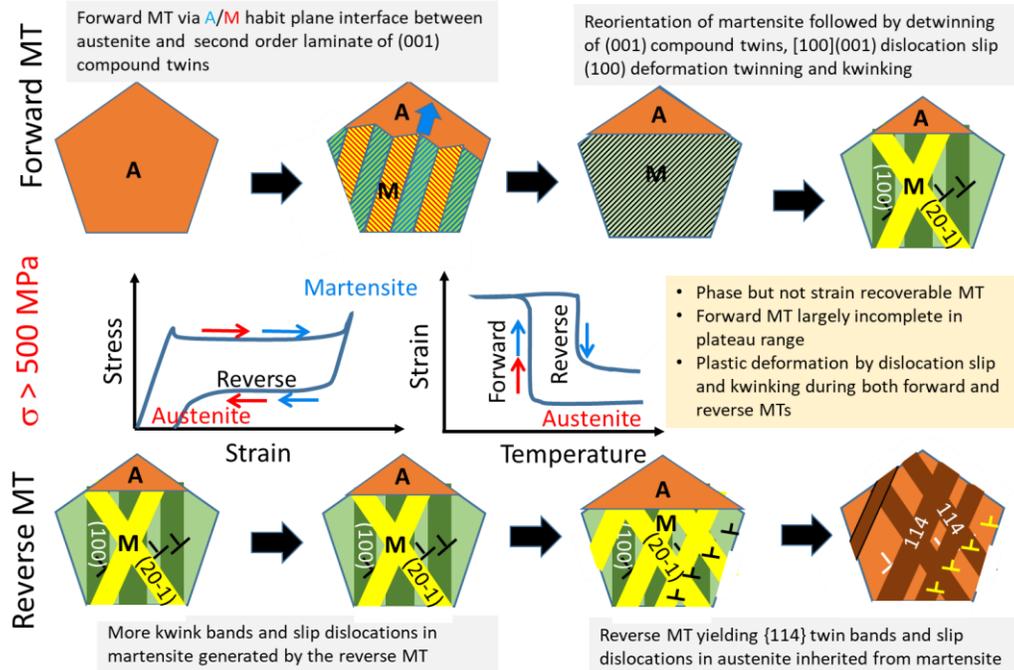

**Figure 14: Forward and reverse MTs under highest stress 800MPa.** Both forward and reverse MTs proceed via a habit plane between austenite and detwinned martensite with the assistance of elastic [31] and plastic deformation (yellow dislocations) in martensite during the forward MT and austenite during the reverse MT. In addition, significant plastic deformation via dislocation slip and kwinking takes place during both forward and reverse MT (black dislocations and green bands turning white and red, respectively, after reverse MT to austenite).

Since the martensite deforms via [100](001) dislocation slip and kwinking during cooling as well as during heating under 800 MPa stress, lattice defects created in martensite are inherited by the austenite during the reverse MT. High dislocation density of slip dislocations and {114} austenite twins (created from {20-1} kwinks [22]) were observed in the microstructure of austenite (Fig. 10g).

It is important to emphasize that we are not suggesting that the significant plastic strains generated by martensitic transformation under stress stem from the requirement for strain compatibility at the habit plane interfaces. When we mention that "MT generates plastic strain," we are referring to the occurrence of plastic deformation during the MT process. If dislocation slip assists in achieving strain compatibility at the habit plane interface, residual slip dislocations would be present in the austenite phase. However, these associated plastic strains do not account for the plastic strains generated by the reverse MTs under stress as evaluated in experiments (Fig. 6).

**4.6 Are the plastic strains and lattice defects generated in austenite and/or in martensite?**

Let us turn attention to the question whether the incremental plastic strains are generated in the austenite and/or martensite. For the forward MT on cooling under external stresses below 500 MPa, no plastic strain (Fig. 5c) and no lattice defects in austenite (Fig. 9b) were observed. This suggests that strain compatibility at habit plane interfaces during the forward MT under stresses below 500 MPa is not assisted by plastic deformation, and the reoriented martensite can withstand external stresses up to 500 MPa. However, when the wire is cooled under stresses higher than 500 MPa (Fig. 5c), the induced martensite deforms via kwinking, as described above

The situation is different during the reverse MT on heating under external stress (Figs. 4,5). In this case, plastic strains are generated at very low external stresses, and their magnitude sharply increases with increasing external stress (Fig. 5d). The magnitudes of plastic strains generated by the reverse MT are higher compared to those generated by the forward MT upon thermal cycling under stress (Fig. 5f). A similar situation occurs during the reverse MT on unloading in isothermal tensile tests (Fig. 7). Since only slip dislocations (no deformation bands) were generated by the reverse MT under 200-500 MPa stress (Fig 10e,f), plastic strains generated by the reverse MT (Fig.5d) are assumed to be due to dislocation slip in austenite and/or martensite.

When the elongated NiTi wire containing reoriented martensite is heated under external stress, it must shorten against the action of the high external stress to recover its parent austenite shape. In thermodynamic terms, this means that the chemical energy stored in the martensite lattice, which is released during the reverse MT upon heating, must overcome the mechanical energy due to external stress without triggering plastic deformation either in the oriented martensite or in austenite. At the same time, there is the requirement for strain compatibility at the habit plane of the reverse MT under stress (Fig. 13). For the case of the 15 ms NiTi #1 wire, reverse MT without plastic deformation is possible only upon heating under stresses lower than ~200 MPa (Figs. 5d, 7b, 10e). When the reverse MT proceeds under external stresses higher than 200 MPa in both isostress (Fig. 4) and isothermal (Fig. 7) tests, we assume the oriented martensite deforms plastically via dislocation slip on the (001) plane prior to and/or during the reverse MT.

On one hand, this is logical since the (001) compound twinned or detwinned martensite filling whole grains of the heated wire is prone to [100](001) dislocation slip [34]. However, as there is only a single [100](001) slip system in the low-symmetry martensite (Fig. S11,S12), and a polycrystalline wire cannot undergo large plastic deformation by the activity of a single slip system, plastic deformation of martensite by [100](001) dislocation slip is constrained equally as is the detwinning of (001) compound twins (Appendix C in [22]). Despite this constraint, plastic strain generated by the reverse MT increases from 0.2% at 200 MPa up to 5% at 500 MPa (Fig. 5d). How this happens is not very clear. One possibility is that dislocation slip in martensite substitutes detwinning of (001) compound twins when the martensite is heated under external stress (Fig. S11,S12), as will be further elaborated in section 4.7. Another, albeit unlikely, possibility is that plastic deformation takes place via dislocation slip in austenite.

The plastic strains generated by the reverse MT upon heating under stress decrease the recoverable strains evaluated in closed loop thermomechanical load cycles (Fig. 5d) so that the wire eventually extends on heating instead of shortening (Fig. 4l). We have investigated reverse MT under extreme constraints in Refs. [33,38]. An interesting change of the structure of the martensite heated under 750 MPa external stress was analyzed in detail by TEM. It was found that the martensite, when exposed to high stress at high temperatures on heating, started to deform plastically by kwinking and displayed modulated monoclinic structure before it transformed to the plastically deformed austenite.

In summary, the experiments (Figs. 4-10) show that plastic deformation generated by both forward and reverse MT occurs either via dislocation slip at low applied stresses and/or kwinking deformation in oriented martensite at high applied stresses. Nevertheless, there are at least two reasons why plastic deformation in austenite shall not be flatly excluded. First, dislocation slip in austenite may be involved in the resolving the strain compatibility at the habit plane interface propagating during the reverse MT (Fig. 13). Second, slip dislocations created within plastically deformed martensite, which are inherited by the austenite during the reverse transformation, may glide in austenite under high external tensile stresses and high temperatures [12-14]. Although one cannot completely exclude plastic deformation in austenite, plastic deformation via dislocation slip in martensite during the reverse MT are proposed to be mainly responsible for cyclic instability and functional fatigue of thermomechanically cycled nanocrystalline NiTi wires (Fig. 8). Of course, this is a prediction which needs to be verified.

**4.7 Does plastic deformation of martensite affect recoverable transformation strains?**

The recoverable transformation strains displayed by NiTi polycrystals are known to depend on the crystallography of MT [1], austenite microstructure [50] and austenite texture [51]. besides of that, there are restrictions on maximum recoverable strains of NiTi polycrystals stemming from the requirement on strain compatibility at grain boundaries in the polycrystals, as discussed by Kohn and Bhattacharya [52]. These restrictions are the reason why the (001) compound twins in grains do not detwin when forward MT proceeds upon cooling under stress (Fig. 4,5) or when the martensite is stress induced in isothermal tests [22,28]. The maximum recoverable strains observable in conventional isothermal and isostress cyclic loading experiments on NiTi polycrystals are restricted by the polycrystalline constraint on detwinning of (001) compound twins. See Appendix C in Ref. [22] for detailed explanation of the effect of intergranular strain compatibility on detwinning (001) compound twins in grains in the 15 ms NiTi #5 wire. The presence of (001) compound twins in martensite variant microstructures in grains created by forward MTs under stress [27,28] explains why maximum recovered strain reached only ~7% in the case of the 15 ms NiTi #1 wire (Fig. 5) and not the theoretically achievable 11% strain of <111> textured NiTi polycrystal [1,50]. The results of tests involving cooling NiTi wires under high stress (Figs. 4,5c) and mechanical loading beyond the end of the stress plateau [28] suggest that plastic deformation of martensite accompanying the forward MT elongates the wire but has a very limited impact on recoverable strains.

Very large recoverable strains were observed in closed loop shape memory tests (Fig. 8g,h) suggesting the possibility of detwinning under high stresses. For example, tensile deformation of a similar 16 ms NiTi #1 wire (grain size 500 nm) at low temperature -90 °C up to 16% strain, followed by unloading and heating above the $A_f$ temperature [53] resulted in recoverable strains exceeding 12% (includes elastic strain). Remarkably, very large recoverable strains ~10% (includes elastic strain) were observed even when this wire was deformed up to ~50 % strain (see Figs. 4,5 in [53]). To verify this suspicion, a closed loop shape memory experiment was performed on the 15 ms NiTi #1 superelastic wire (Fig. S8). Recoverable strains exceeding ~12% (includes elastic strain) were recorded in the test (Fig. S8c,d). If maximum stress in the shape memory cycle is limited to stresses below 900 MPa (Fig, S8a,b), cyclically stable stress-strain-temperature response involving ~10 % recoverable strain is readily obtained (Fig. 8g,h). Such remarkably large recoverable strain can only be explained by considering the detwinning of (001) compound twin laminates in grains of the 15 ms NiTi #1 wire loaded up to high tensile stress, at which plastic deformation of the oriented martensite by kwinking starts. As a result of detwinning at high stresses, the maximum inelastic recoverable strains increased from ~7% in conventional actuator cycles (Figs. 5,7,8) up to ~9% in closed loop shape memory cycle (Fig. S8). This suggests that plastic deformation of martensite by kwinking at low temperature enables the detwinning of (001) compound twins without restricting strain recoverability.

Similarly, the large plastic deformation by kwinking during the forward MT upon cooling under stress (Fig. 5c) does not restrict recoverable strains.

The (001) compound twins in oriented martensite are subjected to high tensile stress also during the reverse MT upon heating under stress (Fig. 4). In that case, the wire deforms plastically in tension, presumably via [100](001) dislocation slip in martensite, which substitutes detwinning of (001) compound twins (Fig. S11,S12), decreasing thus the recoverable transformation strains. However, the extent of this effect largely depends on the constraint applied during the reverse MT upon heating. In this regard, the results of the supplementary thermomechanical loading test involving thermal excursion under constant 8% strain (Fig. S12) are particularly insightful. It is apparent that significant plastic deformation (~4%) occurs when the wire is heated up to 150°C, coinciding with the stress in the wire reaching the kwinking stress (Fig. S12).

Recoverable strains, however, differ among various NiTi wires. Maximum recoverable strains of ~5% generated by the forward MT on cooling the 15 ms NiTi #5 SME wire [20,27,28] are lower than maximum recoverable strain ~7% achievable with the 15 ms NiTi #1 superelastic wire (Fig. 5). This difference is likely attributable to the less sharp <112> fiber texture of the 15 ms NiTi #5 wire (Fig. S9). Additionally, the recoverable strain may potentially be lower due to the lower resistance of the stress-induced martensite to the [100](001) dislocation slip substituting the detwinning in the 15 ms NiTi #5 SME wire. This wire displays ~0.7 % plastic strain when loaded up to the end of the reorientation plateau (Fig. S3) and ~2% plastic strain when loaded up to 12% strain (Fig. S5). The maximum recoverable strains recorded in shape memory tests on the 15 ms NiTi #5 (Fig. S3c,d) are also lower compared to the 15 ms NiTi #1 wire (Fig. S8) due to the less sharp texture (Fig. S9) and/or due to the [100](001) dislocation slip substituting the detwinning of (001) compound twins. Cyclically stable stress-strain-temperature responses in closed loop thermomechanical cycles cannot be achieved with the 15 ms NiTi #5 SME wire due to plastic strains generated upon tensile deformation in the martensite state (Fig. S3,S5,S6).

Strengthening the B19' martensite lattice in NiTi against [100](001) dislocation slip by decreasing grain size [45] or via $Ti_3Ni_4$ nanoprecipitation [43], shows promise in improving the cyclic stability of NiTi, albeit with limited success. A more effective approach involves modifying the transformation pathway by altering the chemical composition which can significantly increase the barrier for [100](001) dislocation slip. For example, orthorhombic martensite in NiTiCu alloys [54] or monoclinic martensite in NiTiHf alloys [55] possess martensite crystal lattices that resist plastic deformation via [100](001) dislocation slip. NiTiHf alloys, renowned for their excellent cyclic stability at high temperatures [55], transform into monoclinic martensite with a markedly different monoclinic lattice. The resistance of NiTiHf alloys to plastic deformation is deemed to be due to H-phase precipitation, which, however, simultaneously suppresses martensite reorientation [56] which is a prerequisite for [100](001) dislocation slip.

## 5. Conclusions

Superelastic 15 ms NiTi #1 wire was subjected to tensile thermomechanical loading tests to determine the stress and temperature thresholds at which plastic deformation occurs. The results showed that the wire is highly resistant to plastic deformation, with a yield stress ~ 900 MPa over a wide temperature range of -100 °C to 400 °C. However, whenever martensitic transformation took place under external stress above certain stress thresholds, incremental plastic strains were recorded. Key findings from the tests include:

1. Forward and reverse martensitic transformations occurring above specific stress thresholds generate characteristic incremental plastic strains.
2. While the forward MT upon cooling under stress does not induce plastic strain up to 500 MPa, the reverse transformation upon heating initiates plastic strains at just 100 MPa stress. The magnitudes of generated plastic strains increase with rising stress levels.
3. Reverse MT upon heating (unloading) generates larger plastic strains than the forward MT upon cooling (loading).
4. Stress-strain-temperature responses in cyclic closed loop shape memory test are stable, with no plastic strains generated despite of ~10 % maximum strains and ~600 MPa maximum stresses.
5. The forward and reverse martensitic transformations generate plastic strains via [100](001) dislocation slip in the martensite phase at low applied stresses and via plastic deformation of martensite by kwinking at high applied stresses.
6. While plastic strains generated by the forward martensitic transformation elongate the wire, those from the reverse martensitic transformation not only elongate the wire but also reduce the recoverable transformation strain.
7. The stress thresholds and magnitudes of plastic strains generated by the forward and reverse transformations represent functional fatigue limits for NiTi. Stable stress-strain-temperature responses in cyclic thermomechanical loading tests are ensured as long as transformations occur below these thresholds.

## Declaration of Competing Interest

The authors declare that they have no known competing financial interests or personal relationships that could have appeared to influence the work reported in this paper.

## Acknowledgments

Support from Czech Science Foundation (CSF) projects 22-15763S (Heller) and 22-20181S (Sittner) is acknowledged. P. Šittner acknowledges support from Czech Academy of Sciences through Praemium Academiae. MEYS of the Czech Republic is acknowledged for the support of infrastructure projects, CNL (CzechNanoLab LM2018110) and Ferrmion (CZ.02.01.01/00/22_008/0004591).

## Data availability

Data will be made available on request.

## References

1. K. Otsuka, X. Ren, Physical metallurgy of Ti–Ni-based shape memory alloys, Progress in Materials Science 50 (2005) 511-678, https://doi.org/10.1016/j.pmatsci.2004.10.001
2. R. Sidharth, A.S.K Mohammed,.H. Sehitoglu, H. Functional Fatigue of NiTi Shape Memory Alloy: Effect of Loading Frequency and Source of Residual Strains. Shap. Mem. Superelasticity **8**, 394–412 (2022). https://doi.org/10.1007/s40830-022-00397-8
3. D. Song, G. Kang, Q. Kan, Ch.Yu and Ch. Zhang, The effect of martensite plasticity on the cyclic deformation of super-elastic NiTi shape memory alloy, Smart Materials and Structures, 23, (2014) 015008, https://doi.org/10.1088/0964-1726/23/1/015008
4. P. Sedmák, P. Šittner, J. Pilch, C. Curfs, Instability of cyclic superelastic deformation of NiTi investigated by synchrotron X-ray, Acta Materialia, 94, (2015), 257-270, https://doi.org/10.1016/j.actamat.2015.04.039Get rights and content
5. O. Tyc, L.Heller, P.Sittner, Lattice defects generated by cyclic thermomechanical loading of superelastic NiTi wire, Shap. Mem. Superelasticity (2021), https://doi.org/10.1007/s40830-021-00315-4
6. J. Hurley, A. M. Ortega, J. Lechniak, K. Gall, H. J. Maier, Structural evolution during the cycling of NiTi shape memory alloys, Z. Metallkd. 94 (2003) 5, https://doi.org/10.1515/ijmr-2003-0096
7. R. Delville, B. Malard, J. Pilch, P. Sittner, D. Schryvers, Transmission electron microscopy investigation of dislocation slip during superelastic cycling of Ni–Ti wires, International Journal of Plasticity 27 (2011) 282–97, https://doi.org/10.1016/J.IJPLAS.2010.05.005
8. O. Molnarova, E. Iaparova, O. Tyc, L. Heller, P. Sittner, TEM analysis of dislocation defects in austenite generated by forward and reverse martensitic transformations in NiTi under stress. 2024, in preparation

9. D. Song, G. Kang, Q.Kan, Ch. Yu, Ch. Zhang, Experimental observations on uniaxial whole-life transformation ratchetting and low-cycle stress fatigue of super-elastic NiTi shape memory alloy micro-tubes *Smart Mater. Struct.* 24(2015)075004 https://doi.org/10.1088/0964-1726/24/7/075004
10. L. Heller, H. Seiner, P. Šittner, P. Sedlák, O. Tyc, L. Kadeřávek, On the plastic deformation accompanying cyclic martensitic transformation in thermomechanically loaded NiTi, International Journal of Plasticity, 111 (2018)53-71, https://doi.org/10.1016/j.ijplas.2018.07.007 .
11. G. Eggeler, E. Hornbogen, A. Yawny, A. Heckmann, M. Wagner, Structural and functional fatigue of NiTi shape memory alloys, Materials Science and Engineering A 378 (2004) 24–33, https://doi.org/10.1016/j.msea.2003.10.327
12. D.M. Norfleet, P.M. Sarosi, S. Manchiraju, M.F.X. Wagner, M.D. Uchic, P.M. Anderson, M.J. Mills, Transformation-induced plasticity during pseudoelastic deformation in Ni-Ti microcrystals, Acta Mater 57 (2009)3549-3561, http://dx.doi.org/10.1016/j.actamat.2009.04.009.
13. H. M. Paranjape, M. L. Bowers, M. J. Mills, P. M. Anderson, Mechanisms for phase transformation induced slip in shape memory alloy micro-crystals, Acta Materialia, 132, 2017, 444-454, https://doi.org/10.1016/j.actamat.2017.04.066
14. M.L. Bowers, X. Chen, M. De Graef, P.M. Anderson, M.J. Mills, Characterization and modeling of defects generated in pseudoelastically deformed NiTi microcrystals Scripta Materialia 78–79, (2014), 69-72, https://doi.org/10.1016/j.scriptamat.2014.02.001
15. Y. Liu, Z.L. Xie, Twinning and detwinning of 〈0 1 1〉 type II twin in shape memory alloy, Acta Materialia, 51(2003)5529-5543, https://doi.org/10.1016/S1359-6454(03)00417-8.
16. K. Knowles, D. Smith, The crystallography of the martensitic transformation in equiatomic nickeltitanium, Acta Metallurgica 29 (1981) 101–110.
17. P. Šittner, P. Sedlák, H. Seiner, P. Sedmák, J. Pilch, R. Delville, L. Heller, L. Kadeřávek, On the coupling between martensitic transformation and plasticity in NiTi: experiments and continuum based modelling, Progress in Materials Science 98 (2018) 249-298, https://doi.org/10.1016/j.pmatsci.2018.07.003
18. P. Šittner, O. Molnárová, L. Kadeřávek, O. Tyc, L. Heller, Deformation twinning in martensite affecting functional behaviour of NiTi shape memory alloys, Materialia 9 (2020)100506, https://doi.org/10.1016/j.mtla.2019.100506
19. S. Manchuraju, A. Kroeger, C. Somsen, A. Dlouhy, G. Eggeler, P.M. Sarosi, P.M. Anderson, M.J. Mills, Pseudoelastic deformation and size effects during in situ transmission electron microscopy tensile testing of NiTi, Acta Materialia, 60(2012)2770-2777, https://doi.org/10.1016/j.actamat.2012.01.043.

20. E. Iaparova, L. Heller, O. Tyc, P. Sittner, Thermally induced reorientation and plastic deformation of B19' monoclinic martensite in nanocrystalline NiTi wires, Acta Materialia,242 (2023) 118477, https://doi.org/10.1016/j.actamat.2022.118477.

21. O. Matsumoto, S. Miyazaki, K. Otsuka, H. Tamura, Crystallography of martensitic transformation in Ti-Ni single crystals, Acta Metallurgica 35 (1987) 2137–2144. https://doi.org/10.1016/0001-6160(87)90042-3

22. O. Molnárová, O. Tyc, L. Heller, H. Seiner, P. Šittner Evolution of martensitic microstructures in nanocrystalline NiTi wires deformed in tension, Acta Materialia 218 (2021) 117166, https://doi.org/10.1016/j.actamat.2021.117166

23. Li Hu, S. Jiang, S. Liu, Y. Zhang, Y. Zhao, Ch. Zhao,Transformation twinning and deformation twinning of NiTi shape memory alloy, Materials Science and Engineering: A, 660(2016) 1-10, https://doi.org/10.1016/j.msea.2016.02.066.

24. X.B.Shi, L.S. Cui, Z.X.Liu, D.G.Jiang, X.D.Han Microstructure of stress-induced martensite in nanocrystalline NiTi shape memory alloy. *Rare Met.* **33**(2014) 379–382, https://doi.org/10.1007/s12598-014-0343-y

25. T. Waitz, The self-accommodated morphology of martensite in nanocrystalline NiTi shape memory alloys Acta Materialia 53 (2005) 2273–2283, https://doi.org/10.1016/j.actamat.2005.01.033

26. X. Bian, L. Heller, O. Tyc, L. Kaderavek, P. Sittner, In-situ synchrotron x-ray diffraction texture analysis of tensile deformation of nanocrystalline superelastic NiTi wire at various temperatures. Mater Sci Eng A 853 (2022) 143725, https://doi.org/10.1016/j.msea.2022.143725

27. O. Tyc, X. Bian, O. Molnárová, L. Kaderavek, L. Heller, P. Šittner, Martensitic transformation induced by cooling NiTi wire under various tensile stresses: martensite variant microstructure, textures, recoverable strains and plastic strains, Applied Materials Today, 2024, submitted

28. O. Tyc, E. Iaparova, O. Molnarova, L. Heller, P. Sittner, Stress induced martensitic transformation in NiTi at elevated temperatures: martensite variant microstructures, recoverable strains and plastic strains, Acta Materialia,2024, submitted

29. C. Cayron, The correspondence theory and its application to niti shape memory alloys, Crystals 12(2022). https://doi.org/10.3390/cryst12020130

30. J.F. Xiao, C. Cayron, and R.E. Logé, Revealing the microstructure evolution of the deformed superelastic NiTi wire by EBSD, Acta Materialia, 255 (2023) 119069, https://doi.org/10.1016/j.actamat.2023.119069

31. L. Heller and P. Sittner, On the Habit Planes Between Elastically Distorted Austenite and Martensite in NiTi, Acta Materialia, 269 (2024) 119828, https://doi.org/10.1016/j.actamat.2024.119828
32. S.Miyazaki, S.Kimura, K,Otsuka and Y.Suzuki, The Habit Plane and Transformation Strains associated with the Martensitic Transformation in Ti-Ni single crystals, Scripta Metallurgica, 18(1984),883-88, https://doi.org/10.1016/0036-9748(84)90254-0
33. P. Šittner, E. Iaparova, L. Kadeřávek, Y. Chen, O. Tyc, Tensile deformation of NiTi shape memory alloy thermally loaded under applied stress, Materials & Design, 226 (2023) 111638, https://doi.org/10.1016/j.matdes.2023.111638
34. T. Ezaz, H. Sehitoglu, H.J. Maier, Energetics of twinning in martensitic NiTi, Acta Materialia,59, (2011) 5893-5904, https://doi.org/10.1016/j.actamat.2011.05.063.
35. P. Chowdhury, H. Sehitoglu, Deformation physics of shape memory alloys – Fundamentals at atomistic frontier, Prog. Mater. Sci. 88, (2017) 49–88, https://doi.org/10.1016/J.PMATSCI.2017.03.003
36. H. Seiner, P. Sedlak, M. Frost, P. Sittner, Kwinking as the plastic forming mechanism of B19' NiTi martensite, Int. Journal of Plasticity 168(2023) 103697, https://doi.org/10.1016/j.ijplas.2023.103697
37. O. Molnárová, M. Klinger, J. Duchoň, H. Seiner, P Šittner, Plastic deformation of B19' monoclinic martensite in NiTi shape memory alloys: HRTEM analysis of interfaces in martensite variant microstructures, Acta Materialia 258, (2023)119242, https://doi.org/10.1016/j.actamat.2023.119242
38. Y Chen, M Klinger, J Duchoň, P Šittner, Modulated martensite in NiTi shape memory alloy exposed to high stress at high temperatures, Acta Materialia, 258(2023) 119250, https://doi.org/10.1016/j.actamat.2023.119250
39. Y. Chen, O. Tyc, O. Molnárová, L. Heller, P. Šittner, Tensile deformation of superelastic NiTi wires in wide temperature and microstructure ranges, Shape Memory and Superelasticity 5 (2019) 42-62, https://doi.org/10.1007/s40830-018-00205-2
40. H. Lin, P. Hua, Q. Sun, Effects of grain size and partial amorphization on elastocaloric cooling performance of nanostructured NiTi, Scripta Materialia,209(2022)114371, https://doi.org/10.1016/j.scriptamat.2021.114371.
41. Ch. Lv, K. Wang, B. Wang, J. Zheng, K. Zhang, G. Li, Y. Lai, Y. Fu, H. Hou, X. Zhao, Coexistence of strain glass transition and martensitic transformation in highly nickel-rich ferroelastic alloy with large elastocaloric effect, Acta Materialia, 264(2024)119598, https://doi.org/10.1016/j.actamat.2023.119598 .

42. Yi-Ting Hsu, Cheng-Tien Wu, Chih-Hsuan Chen, Nanoscale-precipitate-strengthened (Ni,Cu)-rich TiNiCu shape memory alloy with stable superelasticity and elastocaloric performance, Journal of Alloys and Compounds, 997(2024)174937, https://doi.org/10.1016/j.jallcom.2024.174937
43. O. Tyc, O. Molnárová, P. Šittner, Effect of microstructure on fatigue of superelastic NiTi wires, International Journal of Fatigue,152(2021)106400, https://doi.org/10.1016/j.ijfatigue.2021.106400
44. J. Cui, Y. S. Chu, O. O. Famodu, Y. Furuya, J. Hattrick-Simpers, R. D. James, A. Ludwig, S. Thienhaus, M. Wuttig, Z. Zhang, I. Takeuchi. Combinatorial search of thermoelastic shape-memory alloys with extremely small hysteresis width. Nat. Mater. 5, 286–290 (2006). https://doi.org/10.1038/nmat1593
45. P. Hua, M. Xia, Y. Onuki, Y., Q.P. Sun, Nanocomposite NiTi shape memory alloy with high strength and fatigue resistance. Nat. Nanotechnol. **16**, 409–413 (2021). https://doi.org/10.1038/s41565-020-00837-5
46. T. Ezaz, J. Wang, H. Sehitoglu, H.J. Maier, Plastic deformation of NiTi shape memory alloys, Acta Materialia, 61(2013)67-78, https://doi.org/10.1016/j.actamat.2012.09.023.
47. X. Bian, L. Heller, L. Kaderavek, P. Sittner, In-situ synchrotron x-ray diffraction texture analysis of tensile deformation of nanocrystalline NiTi wire in martensite state. Appl Mater Today 26 (2022) 101378, https://doi.org/10.1016/j.apmt.2022.101378
48. P. Sedmák, J. Pilch, L. Heller, J. Kopeček, J. Wright, P. Sedlák P, H. Seiner, P. Sittner, Grain-resolved analysis of localized deformation in nickel-titanium wire under tensile load. Science 2016;353:559–562. https://doi.org/10.1126/science.aad6700
49. A.S.K. Mohammedand H. Sehitoglu, Thermodynamic rationale for transformation-induced slip in shape memory alloys, Acta Materialia, (2024) https://doi.org/10.1016/j.actamat.2024.119998
50. Y. Chen, O. Tyc, L. Kaděřávek, O. Molnárová, L. Heller, P. Šittner, Temperature and microstructure dependence of localized tensile deformation of superelastic NiTi wires, Mater. Des. 174 (2019) 107797, https://doi.org/10.1016/j.matdes.2019.107797
51. H. Inoue, N. Miwa, N. Inakazu, Texture and shape memory strain in TiNi alloy sheets, Acta Materialia, 44 (1996) 4825-4834, https://doi.org/10.1016/S1359-6454(96)00120-6
52. K. Bhattacharya, R.V. Kohn, Symmetry, texture and the recoverable strain of shape-memory polycrystals, Acta Materialia, 44 (1996) 529-542, https://doi.org/10.1016/1359-6454(95)00198-0.
53. Y. Chen, O. Molnárová, O. Tyc, L. Kaděřávek, L. Heller, P. Šittner, Recoverability of large strains and deformation twinning in martensite during tensile deformation of NiTi shape memory alloy polycrystals, Acta Mater.180 (2019) 243–259. https://doi.org/10.1016/j.actamat.2019.09.012

54. C. Chluba, W.W. Ge, R.L. de Miranda, J. Strobel, L.Kienle, E.Quandt, M. Wuttig, Ultralow-fatigue shape memory alloy films. Science 348 (2015)1004–1007, https://dpi.org/10.1126/science.1261164
55. O. Benafan, G.S. Bigelow, A. Garg, R.D. Noebe, D.J. Gaydosh, R.B. Rogers, Processing and scalability of NiTiHf high-temperature shape memory alloys, Shape Memory and Superelasticity 7(2021)109-165, https://doi.org/10.1007/s40830-020-00306-x
56. F. Dalle, E. Perrin, P. Vermaut, M. Masse, R. Portier Interface mobility in Ni49.8Ti42.2Hf8 shape memory alloy, Acta Materialia 50(2002)3557–3565, https://doi.org/10.1016/S1359-6454(02)00151-9